\documentclass[onecolumn, journal,12pt]{ieeeTran}      

\IEEEoverridecommandlockouts                              

\usepackage{color}
\usepackage{psfrag}
\usepackage{graphics} 
\usepackage{anysize}
\usepackage[tbtags]{amsmath} 
\usepackage{amssymb}
\usepackage{verbatim} 
\usepackage{cite}
\usepackage{graphicx}
 \usepackage{epsfig}
\usepackage{psfrag}
\include{spacing}

\newcounter{actr}
{\begin{list}{(\alph{actr})}{\usecounter{actr}}}{\end{list}}

\newcounter{ictr}
{\begin{list}{(\roman{ictr})}{\usecounter{ictr}}}{\end{list}}

\newtheorem{thm}{Theorem}
\newtheorem{lemma}{Lemma}

\newtheorem{corol}{Corollary}

\newtheorem{defn}{Definition}

\newenvironment{new-proof}
{{\em Proof: }}
{ \noindent\qed }

\newcommand{\qed}{\rule[0.1ex]{1.4ex}{1.6ex}}

\newcommand{\defeq}{\stackrel{\Delta}{=}}

\newcommand{\mrm}{\mathrm}

\newcommand{\cC}{{\mathcal{C}}}

\newcommand{\cE}{{\mathcal{E}}}

\newcommand{\cI}{{\mathcal{I}}}

\newcommand{\cJ}{{\mathcal{J}}}

\newcommand{\cM}{{\mathcal{M}}}

\newcommand{\cN}{{\mathcal{N}}}
\newcommand{\CN}{{\mathcal{CN}}}

\newcommand{\cS}{{\mathcal{S}}}

\newcommand{\cT}{{\mathcal{T}}}

\newcommand{\cU}{{\mathcal{U}}}

\newcommand{\cV}{{\mathcal{V}}}

\newcommand{\cX}{{\mathcal{X}}}

\newcommand{\cY}{{\mathcal{Y}}}

\newcommand{\cZ}{{\mathcal{Z}}}

\newcommand{\al}{\alpha}

\newcommand{\G}{\Gamma}

\newcommand{\del}{\delta}

\newcommand{\eps}{\varepsilon}

\DeclareMathAlphabet{\mathbsf}{OT1}{cmss}{bx}{n}
\DeclareMathAlphabet{\mathssf}{OT1}{cmss}{m}{sl}

\DeclareSymbolFont{bsfletters}{OT1}{cmss}{bx}{n}
\DeclareSymbolFont{ssfletters}{OT1}{cmss}{m}{n}
\DeclareMathSymbol{\bsfGamma}{0}{bsfletters}{'000}
\DeclareMathSymbol{\ssfGamma}{0}{ssfletters}{'000}
\DeclareMathSymbol{\bsfDelta}{0}{bsfletters}{'001}
\DeclareMathSymbol{\ssfDelta}{0}{ssfletters}{'001}
\DeclareMathSymbol{\bsfTheta}{0}{bsfletters}{'002}
\DeclareMathSymbol{\ssfTheta}{0}{ssfletters}{'002}
\DeclareMathSymbol{\bsfLambda}{0}{bsfletters}{'003}
\DeclareMathSymbol{\ssfLambda}{0}{ssfletters}{'003}
\DeclareMathSymbol{\bsfXi}{0}{bsfletters}{'004}
\DeclareMathSymbol{\ssfXi}{0}{ssfletters}{'004}
\DeclareMathSymbol{\bsfPi}{0}{bsfletters}{'005}
\DeclareMathSymbol{\ssfPi}{0}{ssfletters}{'005}
\DeclareMathSymbol{\bsfSigma}{0}{bsfletters}{'006}
\DeclareMathSymbol{\ssfSigma}{0}{ssfletters}{'006}
\DeclareMathSymbol{\bsfUpsilon}{0}{bsfletters}{'007}
\DeclareMathSymbol{\ssfUpsilon}{0}{ssfletters}{'007}
\DeclareMathSymbol{\bsfPhi}{0}{bsfletters}{'010}
\DeclareMathSymbol{\ssfPhi}{0}{ssfletters}{'010}
\DeclareMathSymbol{\bsfPsi}{0}{bsfletters}{'011}
\DeclareMathSymbol{\ssfPsi}{0}{ssfletters}{'011}
\DeclareMathSymbol{\bsfOmega}{0}{bsfletters}{'012}
\DeclareMathSymbol{\ssfOmega}{0}{ssfletters}{'012}

\renewcommand{\defeq}{\triangleq}

\newcommand{\rvS}{{\mathssf{S}}}    

\newcommand{\rva}{{\mathssf{a}}}    

\newcommand{\rvb}{{\mathssf{b}}}    

\newcommand{\rvc}{{\mathssf{c}}}    

\newcommand{\rvd}{{\mathssf{d}}}    

\newcommand{\rvk}{{\mathssf{k}}}    
\newcommand{\rvhk}{{\hat{\mathssf{k}}}}    

\newcommand{\rvl}{{\mathssf{l}}}    

\newcommand{\rvm}{{{\mathssf{m}}}}    
\newcommand{\rvmx}{{{\mathssf{m}}}_\mathrm{x}}    

\newcommand{\rvn}{{\mathssf{n}}}    

\newcommand{\rvs}{{\mathssf{s}}}    

\newcommand{\rvt}{{\mathssf{t}}}    

\newcommand{\rvu}{{\mathssf{u}}}    

\newcommand{\rvv}{{\mathssf{v}}}    

\newcommand{\rvw}{{\mathssf{w}}}    

\newcommand{\rvx}{{{\mathssf{x}}}}    

\newcommand{\rvbx}{{\mathbsf{x}}}

\newcommand{\rvy}{{\mathssf{y}}}    

\newcommand{\rvby}{{\mathbsf{y}}}

\newcommand{\rvz}{{\mathssf{z}}}    

\newcommand{\rvht}{{\hat{\rvt}}}

\title{Secret-Key Generation using Correlated Sources and Channels}

\author{Ashish~Khisti,~\IEEEmembership{Student~Member,~IEEE,}
and~Suhas~N.~Diggavi,~\IEEEmembership{Member,~IEEE},
        and~Gregory~W.~Wornell,~\IEEEmembership{Fellow,~IEEE}
        \thanks{Part of the material in this paper was presented
        at the 2008 Information Theory and its Application Workshop~\cite{khisti:08} and the 2008
        International Symposium on Information Theory~\cite{khistiDiggavi:08}. Ashish Khisti was with EECS Department,
        MIT (ashish.khisti@gmail.com).
        Suhas Diggavi is with the faculty of the School of Computer and Communication Sciences at EPFL (suhas.diggavi@epfl.ch).
        Gregory Wornell is with the faculty of EECS Dept., MIT (gww@mit.edu). The  work  of Ashish Khisti and Gregory Wornell was supported in part
        by NSF Grant No. CCF-0515109. The work of Suhas Diggavi was supported in part by the Swiss National Science Foundation through NCCR-MICS
        }  }

\begin{document}
\maketitle

\begin{abstract}
We study the problem of generating a shared secret key between two
terminals in a joint source-channel setup --- the sender
communicates to the receiver over a discrete memoryless wiretap
channel and additionally the terminals have access to correlated
discrete memoryless source sequences. We establish lower and upper
bounds on the secret-key capacity. These bounds coincide,
establishing the capacity, when the underlying channel consists of
independent, parallel and reversely degraded wiretap channels. In
the lower bound, the equivocation terms of the source and channel
components are functionally additive. The secret-key rate is
maximized by optimally balancing the the source and channel
contributions. This tradeoff is illustrated in detail for the
Gaussian case where it is also shown that Gaussian codebooks achieve
the capacity. When the eavesdropper also observes a source sequence,
the secret-key capacity is established when  the sources and
channels of the eavesdropper are a degraded version of the
legitimate receiver. Finally the case when the terminals also have
access to a public discussion channel is studied. We propose
generating separate keys from the source and channel components and
establish the optimality of this approach when the when the channel
outputs of the receiver and the eavesdropper are conditionally
independent given the input.
\end{abstract}

\section{Introduction}

\label{sec:Introduction}

Many applications in cryptography require that the legitimate
terminals have  shared secret-keys, not available to unauthorized
parties. Information theoretic security encompasses the study of
source and channel coding techniques to  generate secret-keys
between legitimate terminals. In the channel coding literature, an
early work in this area is the wiretap channel
model~\cite{wyner:75Wiretap}. It consists of three terminals --- one
sender, one receiver and one eavesdropper. The sender communicates
to the receiver and the eavesdropper over a discrete-memoryless
broadcast channel. A notion of equivocation-rate --- the normalized
conditional entropy of the transmitted message given the observation
at the eavesdropper, is introduced, and the tradeoff between
information rate and equivocation rate is studied.  Perfect secrecy
capacity, defined as the maximum information rate under the
constraint that  the equivocation rate approaches the information
rate asymptotically in the block length is of particular interest.
Information transmitted at this rate can be naturally used as a
shared secret-key between the sender and the receiver.

In the source coding setup~\cite{ahlswedeCsiszar:93,maurer:93}, the
two terminals observe correlated source sequences and use a public
discussion channel for communication. Any information sent over this
channel is available to an eavesdropper. The terminals generate a
common secret-key that is concealed from the eavesdropper in the
same sense as the wiretap channel --- the equivocation rate
asymptotically equals the secret-key rate. Several multiuser
extensions of this problem have been subsequently studied. See e.g.,
~\cite{csiszarNarayan:00,csiszarNarayan:04}.

Motivated by the above works, we study a problem where the
legitimate terminals observe correlated source sequences and
communicate over a wiretap channel and are required to generate a
common secret-key. One application of this setup is sensor networks,
where terminals measure correlated physical processes. It is natural
to investigate how these measurements can be used for secrecy. In
addition, the sensor nodes communicate over a wireless channel where
an eavesdropper could hear transmission albeit through a different
channel. Another application is secret key generation using
biometric measurements~\cite{draperKhisti:07}. During the
registration phase, an enrollment biometric is stored into a
database. To generate a secret key subsequently, the user is
required to provide another measurement of the same biometric. This
new measurement differs from the enrollment biometric due to factors
such as measurement noise and hence can be modeled as a correlated
signal. Again when the database is remotely located, the
communication happens over a channel which could be wiretapped.

The secret-key agreement scheme,~\cite{maurer:93,ahlswedeCsiszar:93}, generates a secret key only using the source sequences. On the other hand, the wiretap coding scheme~\cite{wyner:75Wiretap} generates a secret-key by exploiting the structure of the underlying broadcast channel.   Clearly in the present setup, we should consider schemes that take into account both the source and channel contributions. One simple approach  is timesharing --- for a certain fraction of time the wiretap channel is used as a (rate limited) transmission channel whereas for the remaining time, a wiretap code is used to transmit information at the secrecy capacity. However such an approach in general is sub-optimal. As we will see, a better approach involves  simultaneously exploiting both the source and channel uncertainties at the eavesdropper.  As our main result we present lower and upper bounds on the secret-key capacity. The lower bound is developed by providing a coding theorem that consists of a combination of a Wyner-Ziv codebook, a wiretap codebook and a secret-key generation codebook.  Our upper and lower bounds coincide, establishing the secret-key-capacity, when the wiretap channel consists of parallel independent and  degraded channels.

We also study the case when the eavesdropper observes a source sequence correlated with the legitimate terminals. The secret-key capacity is established when the  sources sequence of the eavesdropper  is a degraded version of the sequence of the legitimate receiver and the channel of the eavesdropper is a degraded version of the channel of the legitimate receiver. Another variation --- when a public discussion channel is available for interactive communication, is also discussed and the secret-key capacity is established when the channel output symbols of the legitimate receiver and eavesdropper are conditionally independent given the input.

The problem studied in this paper also provides an operational significance for the rate-equivocation region of the wiretap channel. Recall that the
rate-equivocation region captures the tradeoff between the
conflicting requirements of maximizing the information rate to the legitimate receiver and the equivocation level at the eavesdropper~\cite{csiszarKorner:78}. To maximize the contribution of the correlated sources, we must operate at the Shannon capacity of the underlying channel. In contrast, to maximize the contribution of the wiretap channel, we operate at a point of maximum equivocation. In general, the optimal operating point lies in between these extremes. We illustrate this tradeoff in detail for the case of Gaussian sources and channels.

In related work~\cite{merhav06,yamamoto:97,Gunduz:08} study a setup involving sources and channels, but require that a source sequence be reproduced at
 the destination subjected to an equivocation level at the eavesdropper. In contrast our paper does not impose any requirement on reproduction of a source sequence, but instead requires that the terminals generate a common secret key. A recent work, \cite{prabhakaranRamchandran07},  considers transmitting an independent confidential message using correlated sources and noisy channels. This problem is different from the secret-key generation problem, since the secret-key, by definition, is an arbitrary function of the source sequence, while the message is required to be independent of the source sequences.  Independently
 and concurrently of our work the authors of~\cite{prabhakaranEswaran:08}
 consider the scenario of joint secret-message-transmission and secret-key-generation, which when specialized to the case of no secret-message reduces to the scenario treated in this paper. While the expression for the achievable rate in~\cite{prabhakaranEswaran:08} appears consistent with the expression in this paper, the  optimality claims in~\cite{prabhakaranEswaran:08}
 are limited to the case when either the sources or the channel do not provide any secrecy.

The rest of the paper is organized as follows. The problem of interest is formally introduced in section~\ref{sec:Defn} and the main results of this work are summarized in section~\ref{sec:results}. Proofs of the lower and upper bound appear in sections~\ref{sec:AchievMain} and~\ref{sec:UpperBound} respectively. The secrecy capacity for the case of independent parallel reversely degraded channels is provided in section~\ref{sec:ReverselyDegraded}. The case when the wiretapper has access to a degraded source and observes transmission through a degraded channel is treated in section~\ref{sec:WSI} while section~\ref{sec:Discussion} considers the case when a public discussion channel allows interactive communication between the sender and the receiver. The conclusions appear in section~\ref{sec:concl}.
\section{Problem Statement}
\label{sec:Defn}
\begin{figure}[htb]
 \centering \psfrag{U}{$\rvu^N$}\psfrag{V}{$\rvv^N$}
\psfrag{ch}{$p_{\rvy,\rvz |
\rvx}(\cdot,\cdot|\cdot)$}\psfrag{X}{$\rvx^n$}\psfrag{Y}{$\rvy^n$}
\psfrag{Z}{$\rvz^n$}
\includegraphics[scale=0.5]{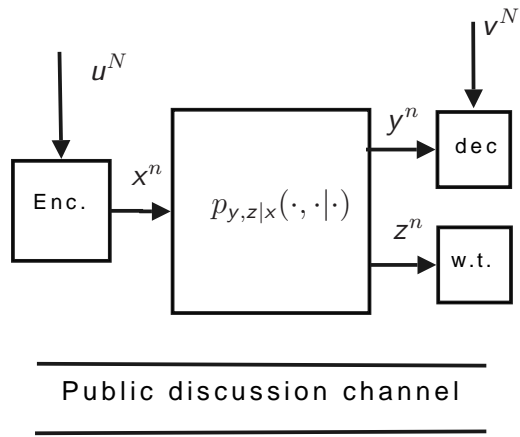}
\caption{Secret-key agreement over the wiretap channel with correlated sources. The sender and receiver communicate over a wiretap channel and have access to correlated sources. They communicate interactively over a public discussion channel of rate $R$, if it is available.}
\label{fig:ch}
\end{figure}

Fig.~\ref{fig:ch} shows the setup of interest. The sender and receiver communicate over a wiretap channel and have access to correlated sources. They can interact over a public-discussion channel. We consider two extreme scenarios: (a) the discussion channel does not exist (b) the discussion channel has unlimited capacity.

The channel from sender to receiver and wiretapper is a discrete-memoryless-channel (DMC), $p_{\rvy,\rvz|\rvx}(\cdot,\cdot|\cdot)$.  The
sender and intended receiver observe discrete-memoryless-multiple-source (DMMS) $p_{\rvu,\rvv}(\cdot,\cdot)$ of length
$N$ and communicate over $n$ uses of the DMC. We separately consider the cases when no public discussion is allowed and unlimited discussion is allowed.

\subsection{No discussion channel is available}
An $(n,N)$ secrecy code is defined as follows. The sender  samples a
random variable $\rvmx$  \footnote{The alphabets associated with
random variables will be denoted by calligraphy letters. Random
variables are denoted by sans-serif font, while their realizations
are denoted by standard font. A length $n$ sequence is denoted by
$x^n$.} from the conditional distribution $p_{\rvm_\rvx
|\rvu^N}(\cdot|u^n)$. The encoding function $f_n: \cM_x \times
{\cU}^N \rightarrow \cX^n$ maps the observed source sequence to the
channel output. In addition, two key generation functions $\rvk =
K_n(\cM_x, \cU^N)$ and $\rvl = L_n(\cV^N, \cY^n)$ at the sender and
the receiver are used for secret-key generation.  A secret-key rate
$R$ is achievable with bandwidth expansion factor $\beta$ if there
exists a sequence of $(n,\beta n)$  codes, such that for a sequence
$\eps_n$ that approaches zero as $n\rightarrow\infty$, we have (i)
$\Pr(\rvk\neq \rvl)\le \eps_n$ (ii) $\frac{1}{n}H(\rvk) \ge
R-\eps_n$ (iii)$\frac{1}{n}I(\rvk;\rvz^n) \le \eps_n$. The
secret-key-capacity is the supremum of all achievable rates.

For some of our results, we will also consider the case when the wiretapper observes a side information sequence $\rvw^N$ sampled i.i.d. $p_\rvw(\cdot)$.  In this case, the secrecy condition in (iii) above is replaced with
\begin{equation}
\frac{1}{n}I(\rvk;\rvz^n,\rvw^N) \le \eps_n\label{eq:EquivWSI}
\end{equation}

In addition, for some of our results we will consider the special case when the wiretap channel
consists of parallel and independent channels each of which is degraded.

\subsubsection{Parallel Channels}

\begin{defn}
\label{defn:ProductBroadcastChannels} A \emph{product} broadcast
channel is one in which the $M$ constituent subchannels have finite
input and output alphabets, are memoryless and independent of each
other, and are characterized by their transition probabilities
\begin{equation}
\Pr\left(\{y_m^n, z_m^n\}_{m=1,\ldots,M} \mid \{x_{m}^n\}_{m=1,\ldots, M}
\right)
= \prod_{m=1}^M \prod_{t=1}^n
\Pr(y_{m}(t),z_{m}(t) \mid x_{m}(t)),
\label{eq:ChanIndep}
\end{equation}
where $x_m^n = (x_m(1), x_m(2),\dots, x_m(n))$ denotes the sequence of
symbols transmitted on subchannel $m$, where
$y_{m}^n = (y_m(1),y_m(2),\ldots, y_m(n))$ denotes the sequence
of symbols obtained by the legitimate receiver on subchannel $m$, and where
$z_{m}^n=(z_{m}(1),z_{m}(2),\ldots,z_{m}(n))$
denotes the sequence of symbols received by the eavesdropper on
subchannel $m$.
\end{defn}
\hfill$\blacksquare$

A special class of product broadcast channels, known as the
reversely degraded broadcast channel~\cite{elGamal:80} are defined as follows.
\begin{defn}
A product broadcast channel is \emph{reversely-degraded} when each of
the $M$ constituent subchannels is degraded in a prescribed order. In
particular, for each subchannel $m$,  one of  $\rvx_m \rightarrow \rvy_m\rightarrow \rvz_m$ or $\rvx_m \rightarrow \rvz_m \rightarrow \rvy_m$ holds.
\label{defn:reverselyDegradedBC}
\end{defn}
\hfill$\blacksquare$

Note that in Def.~\ref{defn:reverselyDegradedBC} the order of degradation need not be the same for all subchannels, so the overall channel need
not be degraded.  We also emphasize that in any subchannel the  receiver and eavesdropper are \emph{physically} degraded.  Our capacity results, however, only depend on the marginal distribution of receivers in each subchannel\footnote{However, when we consider the presence of a public-discussion channel and interactive communication, the capacity does depend on joint distribution $p_{\rvy,\rvz|\rvx}(\cdot)$}.  Accordingly, our results in fact hold for the larger class of channels in which there is only stochastic degradation in the subchannels.

We obtain further results when the channel is Gaussian.

\subsubsection{Parallel Gaussian Channels and Gaussian Sources}
\label{subsec:Gauss}
\begin{defn}
\label{defn:gauss}
A reversely-degraded product broadcast channel is \emph{Gaussian} when
it takes the form
\begin{equation}
\begin{aligned}
\rvy_{m} &= \rvx_{m} + \rvn_{\mrm{r},m},\\
\rvz_{m} &= \rvx_{m}+\rvn_{\mrm{e},m},
\end{aligned}\quad m=1,\dots,M
\label{eq:GaussCommonMsgModel}
\end{equation}
where the noise variables are all mutually independent, and $\rvn_{\mrm{r},m}
\sim \CN(0,\sigma_{\mrm{r},m}^2)$ and $\rvn_{\mrm{e},m}\sim
\CN(0,\sigma_{\mrm{e},m}^2)$.  For this channel, there is also an
average power constraint
\begin{equation*}
E\left[\sum_{m=1}^M \rvx_m^2\right]\le P.
\end{equation*}
\end{defn}
\hfill$\blacksquare$

Furthermore we assume that $\rvu$ and $\rvv$ are jointly Gaussian
(scalar valued) random variables, and without loss of generality we
assume that $\rvu\sim \cN(0,1)$ and $\rvv = \rvu + \rvs$, where $\rvs
\sim \cN(0,S)$ is independent of $\rvu$.

\subsection{Presence of a public discussion channel}
\label{subsec:discussion_channel}
We will also  consider a variation on the original setup when a public discussion channel is  available for communication. This setup was first introduced in the pioneering works~\cite{maurer:93,ahlswedeCsiszar:93} where the secret-key capacity was bounded for source and channel models. The sender and receiver can interactively exchange messages on the public discussion channel.

The sender  transmits symbols $\rvx_1,  \ldots \rvx_n$ at times $0 < i_1 < i_2 < \ldots < i_n$ over the wiretap channel. At these times the receiver and the eavesdropper observe symbols $\rvy_1, \rvy_2, \ldots, \rvy_n$ and $\rvz_1, \rvz_2,\ldots,\rvz_n$ respectively. In the remaining times the sender and receiver exchange messages $\phi_t$ and $\psi_t$ where $1 \le t \le k$. For convenience we let $i_{n+1} = k+1$. The eavesdropper observes both $\phi_t$ and $\psi_t$.
More formally,

\begin{itemize}
\item  At time $0$ the sender and receiver sample
random variables $\rvmx$ and  $\rvm_\rvy$ respectively  from conditional distributions $p_{\rvmx|\rvu^N}(\cdot|u^N)$ and $p_{\rvm_\rvy|\rvv^N}(\cdot|v^N)$.
Note that $\rvmx \rightarrow \rvu^N \rightarrow \rvv^N \rightarrow \rvm_\rvy$ holds.

\item At times $0 < t < i_1$ the sender generates $\phi_t= \Phi_t(\rvmx, \rvu^N, \psi^{t-1})$ and the receiver generates $\psi_t = \Psi_t(\rvm_\rvy, \rvv^N, \phi^{t-1})$. These messages are exchanged over the public channel.

\item  At times $i_j$, $1 \le j \le n$, the sender generates $\rvx_j = X_j(\rvmx, \rvu^N, \psi^{i_j-1})$ and sends it over the channel. The receiver and eavesdropper observe $\rvy_j$ ad $\rvz_j$ respectively. For these times we  set $\phi_{i_j} = \psi_{i_j}= 0$.

\item For times $i_j < t < i_{j+1}$, where $1 \le j \le n$, the sender and receiver compute $\phi_t = \Phi_t(\rvmx, \rvu^N, \psi^{t-1})$ and $\psi_t = \Psi_t(\rvm_\rvy, \rvv^N, \rvy^j, \phi^{t-1})$ respectively and exchange them over the public channel.

\item At time $k+1$, the sender and receiver compute  $\rvk = K_n(\rvmx, \rvu^N, \psi^k) $ and the receiver computes $\rvl =L_n(\rvm_\rvy,\rvv^N, \rvy^n,\phi^k)$.

\end{itemize}

We require that for some sequence $\eps_n$ that vanishes as $n \rightarrow \infty$, $\Pr(\rvk\neq \rvl) \le \eps_n$  and
\begin{equation}
\frac{1}{n} I(\rvk; \rvz^n, \psi^k,\phi^k)\le \eps_n.
\end{equation}

\section{Statement of Main Results}
\label{sec:results}
It is convenient to define the following quantities which will be used in the sequel. Suppose that $\rvt$ is a random variable such
that $\rvt \rightarrow \rvu \rightarrow \rvv $, and $\rva$ and
$\rvb$ are random variables such that $\rvb\rightarrow\rva
\rightarrow \rvx \rightarrow (\rvy,\rvz)$ holds and $I(\rvy;\rvb)\le
I(\rvz;\rvb)$.  Furthermore define
\begin{subequations}
\begin{align}
&R_\mrm{ch} = I(\rva;\rvy), \label{eq:Rch}\\ &R_\mrm{eq}^- = I(\rva;\rvy|\rvb)-I(\rva;\rvz|\rvb)\label{eq:Req}\\
&R_\mrm{s} = I(\rvt ;\rvv),\label{eq:Rs}\\ &R_\mrm{wz} =
I(\rvt;\rvu)-I(\rvt;\rvv).\label{eq:Rwz} \\
&R^+_\mrm{eq} = I(\rvx;\rvy\mid \rvz). \label{eq:Req+}\\
&R_\mrm{ch}^+ = I(\rvx;\rvy), \label{eq:Rchs}
\end{align}\end{subequations}

We establish the following lower and upper bounds on the secret key rate in Section~\ref{sec:AchievMain} and~\ref{sec:UpperBound} respectively.

\begin{lemma}
\label{lem:achievRate}
A lower bound on the secret-key rate is given by
\begin{align}
R^-_\mrm{key} = \beta R_\mrm{s} + R_\mrm{eq}^-,\label{eq:RLB}
\end{align}
where the random variables $\rvt,\rva$ and $\rvb$ defined above additionally
satisfy the condition
\begin{align}
\beta R_\mrm{wz} \le R_\mrm{ch}\label{eq:achievIneq}
\end{align} and the quantities $R_\mrm{wz}$, $R_\mrm{s}$, $R_\mrm{eq}^-$ and $R_\mrm{ch}$ are defined in~\eqref{eq:Rwz},~\eqref{eq:Rs},~\eqref{eq:Req} and~\eqref{eq:Rch} respectively.
\end{lemma}
\hfill$\blacksquare$

\begin{lemma}
\label{lem:upperBound}
 An upper bound on the secret-key rate is given by,
\begin{align}
\label{eq:RUB} R_\mrm{key}^+ = \sup_{\{(\rvx,\rvt)\}}\left\{\beta
R_\mrm{s} + R_\mrm{eq}^+\right\},
\end{align}
where the supremum is over all distributions over the random
variables $(\rvx,\rvt)$ that satisfy $\rvt \rightarrow \rvu
\rightarrow \rvv$, the cardinality of $\rvt$ is at-most the
cardinality of $\rvu$ plus one, and
\begin{align}\label{eq:RubIneq}   \beta
R_\mrm{wz} \le R_\mrm{ch}^+.\end{align} The quantities $R_\mrm{s}$, $R_\mrm{wz}$, $R_\mrm{eq}^+$ and $R_\mrm{ch}^+$
are defined in~\eqref{eq:Rs},~\eqref{eq:Rwz},~\eqref{eq:Req+} and~\eqref{eq:Rchs} respectively.

Furthermore, it suffices to consider only those distributions where
$(\rvx,\rvt)$ are independent.
\end{lemma}
\hfill$\blacksquare$

\subsection{Reversely degraded parallel independent channels}
The bounds in Lemmas~\ref{lem:achievRate} and~\ref{lem:upperBound} coincide for the case of reversely degraded channels as shown in section~\ref{subsec:ParChanProof} and stated in the following theorem.
\begin{thm}
The secret-key-capacity for the reversely degraded parallel independent channels in Def.~\ref{defn:reverselyDegradedBC} is given
by \begin{equation}C_\mrm{key} =
\max_{\{(\rvx_1,\ldots,\rvx_M,\rvt)\}}\left\{\beta I(\rvv;\rvt) +
\sum_{i=1}^M
I(\rvx_i;\rvy_i|\rvz_i)\right\},\label{Eq:RevDegCap}\end{equation}
where the random variables $(\rvx_1,\ldots,\rvx_M,\rvt)$ are
mutually independent, $\rvt \rightarrow \rvu\rightarrow \rvv$,
and
\begin{equation}
\sum_{i=1}^M I(\rvx_i;\rvy_i) \ge
\beta\{I(\rvu;\rvt)-I(\rvv;\rvt)\}\label{eq:RevDegRate}
\end{equation}
\label{corol:RevDegCap} Furthermore, the cardinality of $\rvt$ obeys
the same  bounds as in Lemma~\ref{lem:upperBound}.
\end{thm}
\hfill$\blacksquare$

\subsection{Gaussian Channels and Sources}

\begin{figure}
\centering
\psfrag{x1}{$\rvx_1$}\psfrag{x2}{$\rvx_2$}\psfrag{yr1}{$\rvy_1$}\psfrag{yr2}{$\rvz_2$}
\psfrag{ye1}{$\rvz_1$}\psfrag{ye2}{$\rvy_2$}\psfrag{nr1}{$\rvn_\mrm{r,1}$}\psfrag{nr2}{$\rvn_\mrm{r,2}-\rvn_\mrm{e,2}$}
\psfrag{ne1}{$\rvn_\mrm{e,1} - \rvn_\mrm{r,1}$}\psfrag{ne2}{$\rvn_\mrm{e,2}$}
\includegraphics[scale=0.5]{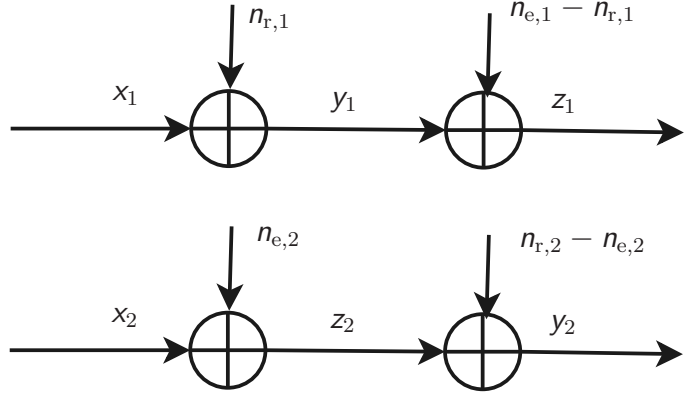}
\caption{An example of independent parallel and reversely degraded Gaussian channels. On the first channel, the eavesdropper channel is noisier than the legitimate receiver's channel while on the second channel the order of degradation is reversed.  }
\label{fig:gauss}
\end{figure}

For the case of Gaussian sources and Gaussian channels, the secret-key capacity can be achieved by Gaussian codebooks as established in section~\ref{subsec:ParChanGaussProof} and stated below.

\begin{corol}
The secret-key capacity for the case of Gaussian parallel channels
and Gaussian sources in subsection~\ref{subsec:Gauss} is obtained by
optimizing~\eqref{Eq:RevDegCap} and~\eqref{eq:RevDegRate} over
independent Gaussian distributions i.e., by selecting $\rvx_i \sim
\cN(0,P_i)$ and $\rvu = \rvt + \rvd$, for some $\rvd \sim \cN(0,D)$,
independent of $\rvt$ and $\sum_{i=1}^n  P_i \le P$, $P_i \ge 0$,
and $0< D  \le 1$.
\begin{equation}
C_\mrm{key}^G =\max_{\{P_i\}_{i=1}^M, D}
\left\{\frac{\beta}{2}\log\left(\frac{1 + S}{D + S}\right) +
\mathop{\sum_{{i: 1\le i \le M}}}_{\sigma_{r,i}\le
\sigma_{e,i}}\frac{1}{2}\log\left(\frac{1 + P_i/\sigma_{r,i}^2}{1 +
P_i/\sigma_{e,i}^2}\right)\right\},
\end{equation}
where $D,P_1,\ldots, P_M$ also satisfy the following relation:
\begin{equation}
\sum_{i=1}^M\frac{1}{2}\log\left(1
+\frac{P_i}{\sigma_{r,i}^2} \right)\ge
\beta\left\{\frac{1}{2}\log\left(\frac{1}{D}\right)-\frac{1}{2}\log\left(\frac{1+S}{D+S}\right)\right\}
\end{equation}
\label{corol:RevDegCapGauss}
\end{corol}
\hfill$\blacksquare$
\subsection{Remarks}

\begin{enumerate}

\item
Note that the secret-key capacity expression~\eqref{Eq:RevDegCap}
exploits both the source and channel uncertainties at the
wiretapper. By setting either uncertainty to zero, one can recover
known results. When $I(\rvu;\rvv)=0$, i.e., there is no secrecy from
the source, the secret-key-rate equals the wiretap
capacity~\cite{wyner:75Wiretap}. If  $I(\rvx;\rvy|\rvz)=0$, i.e.,
there is no secrecy from the channel, then our result essentially
reduces to the result by Csiszar and
Narayan~\cite{csiszarNarayan:00}, that consider the case when the
channel is a noiseless bit-pipe with finite rate.

\item
In general, the setup of wiretap channel involves a tradeoff between
information rate and equivocation. The secret-key generation setup
provides an operational significance to this tradeoff. Note that the
capacity expression~\eqref{Eq:RevDegCap} in
Theorem~\ref{corol:RevDegCap} involves two terms. The first term
$\beta I(\rvt;\rvv)$ is the contribution from the correlated
sources. In general, this quantity increases by increasing the
information rate $I(\rvx;\rvy)$ as seen from~\eqref{eq:RevDegRate}.
The second term, $I(\rvx;\rvy|\rvz)$ is the equivocation term and
increasing this term, often comes at the expense of the information
rate. Maximizing the secret-key rate, involves operating on a
certain intermediate point on the rate-equivocation tradeoff curve as illustrated by an example below.

Consider a pair of Gaussian parallel channels,

\begin{equation}
\label{eq:parChan}
\begin{aligned}
\rvy_{1} &= a_1 \rvx + \rvn_{r,1},\quad \rvz_{1} = b_1 \rvx + \rvn_{e,1} \\
\rvy_{2} &= a_2 \rvx + \rvn_{r,2},\quad \rvz_{2} = \rvy_{2}
\end{aligned}\end{equation}
where $a_1 = 1$, $a_2 = 2$, and $b_1 = 0.5$. Furthermore,  $\rvu
\sim \cN(0,1)$ and $\rvv = \rvu + \rvs$, where $\rvs \sim \cN(0,1)$
is independent of $\rvu$. The noise variables are all sampled from the $\CN(0,1) $ distribution and
appropriately correlated so that the users are degraded on each
channel. A total power constraint $P=1$ is selected and the
bandwidth expansion factor $\beta$ equals unity.

From Theorem~\ref{corol:RevDegCapGauss},
\begin{align}
&C_\mrm{key} = \max_{P_1, P_2, D} R_\mrm{eq}(P_1,P_2) + \frac{1}{2}\log\frac{2}{1 + D} ,\label{eq:ExampleCap}\\
& \text{such that},\notag\\
&R_\mrm{wz}(D) = \frac{1}{2}\log\frac{1}{D}-
\frac{1}{2}\log\frac{2}{1 + D} \label{eq:ExampleDist}\\
&\quad\quad\le \frac{1}{2}\left(\log\left(1 + a_1^2 P_1 \right) +
\log(1+ a_2^2
P_2)\right),\\
&R_{\mrm{eq}}(P_1,P_2) = \frac{1}{2}\left(\log(1+ a_1^2 P_1) -
\log(1 + b_1^2 P_1)\right).\label{eq:ExampleEquiv}
\end{align}

\begin{figure}
\centering
\includegraphics[scale=0.5]{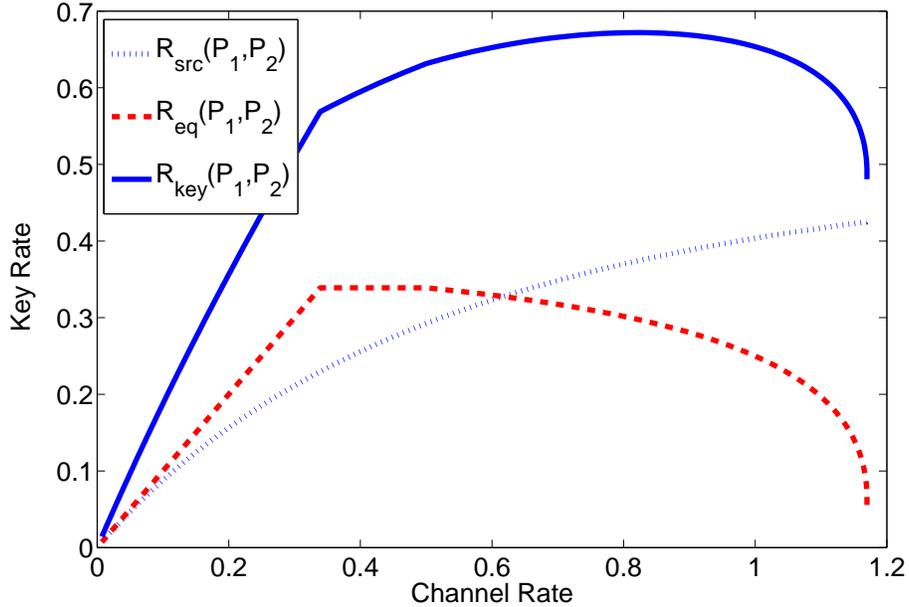}
\caption{Tradeoff inherent in the secret-key-capacity formulation.
The solid curve is the secret-key-rate, which is the sum of the two
other curves. The dotted curve represents the source equivocation,
while the dashed curve represents the channel equivocation
~\eqref{eq:ExampleEquiv}. The secret-key-capacity is
obtained at a point between the maximum equivocation and maximum
rate. } \label{fig:req}

\vspace{-2em}
\end{figure}

Fig.~\ref{fig:req} illustrates the (fundamental) tradeoff between
rate and equivocation for this channel, which is obtained as we vary
power allocation between the two sub-channels. We also present the
function $R_\mrm{src}= I(\rvt;\rvv)$ which monotonically increases
with the rate, since larger the rate, smaller is the distortion in
the source quantization. The optimal point of operation is between
the point of maximum equivocation and maximum rate as indicated by
the maximum of the solid line in Fig.~\ref{fig:req}. This
corresponds to a power allocation $(P_1,P_2) \approx (0.29,0.71)$
and the maximum value is $R_\mrm{key}\approx 0.6719$.

\end{enumerate}

\subsection{Side information at the wiretapper}

So far, we have focussed on the case when there is no side
information at the wiretapper. This assumption is valid for certain
application such as biometrics, when the correlated sources
constitute successive measurements of a person's biometric. In other
applications, such as sensor networks, it is more realistic to
assume that the wiretapper also has access to a side information
sequence.

We consider the setup described in Fig.~\ref{fig:ch}, but with a
modification that the wiretapper observes a source sequence $\rvw^N$,
obtained by $N-$ independent samples of a random variable $\rvw$. In
this case the secrecy condition takes the form in~\eqref{eq:EquivWSI}. We only consider the case when the sources and channels satisfy a degradedness condition.

\begin{thm}
\label{lem:WSI} Suppose that the random variables $(\rvu,\rvv,\rvw)$
satisfy the degradedness condition
$\rvu\rightarrow\rvv\rightarrow\rvw$ and the broadcast channel is
also degraded i.e., $\rvx\rightarrow\rvy\rightarrow\rvz$. Then, the
secret-key-capacity is given by
\begin{equation}
C_\mrm{key} = \max_{(\rvx,\rvt)}\left\{\beta
(I(\rvt;\rvv)-I(\rvt;\rvw)) + I(\rvx;\rvy|\rvz)
\right\},\label{eq:wiretapSICap}
\end{equation}
where the maximization is over all random variables $(\rvt,\rvx)$
that are mutually independent, $\rvt\rightarrow \rvu\rightarrow
\rvv\rightarrow \rvw$ and
\begin{equation}
I(\rvx;\rvy) \ge
\beta(I(\rvu;\rvt)-I(\rvv;\rvt))\label{eq:wiretapSIRate}
\end{equation}
holds. Furthermore, it suffices to optimize over random variables
$\rvt$ whose cardinality does not exceed that of $\rvu$ plus two.
\end{thm}
\hfill$\blacksquare$
\subsection{Secret-key capacity with a public discussion channel}

When public interactive communication is allowed as described in section~\ref{subsec:discussion_channel}, we have the following upper bound on the secret-key capacity.

\begin{thm}
An upper bound on the secret-key capacity for source-channel setup
with a public discussion channel is
\begin{equation}
C_\mrm{key}\le \max_{p_\rvx} I(\rvx;\rvy|\rvz) + \beta I(\rvu;\rvv).
\label{eq:Ckey_PubDisc}
\end{equation}

The upper bound is tight when channel satisfies either $\rvx \rightarrow \rvy \rightarrow \rvz$ or $\rvy\rightarrow \rvx\rightarrow \rvz$.
\label{thm:secInteractive}
\end{thm}
\hfill$\blacksquare$
\begin{figure}
\centering
\includegraphics[scale=0.5]{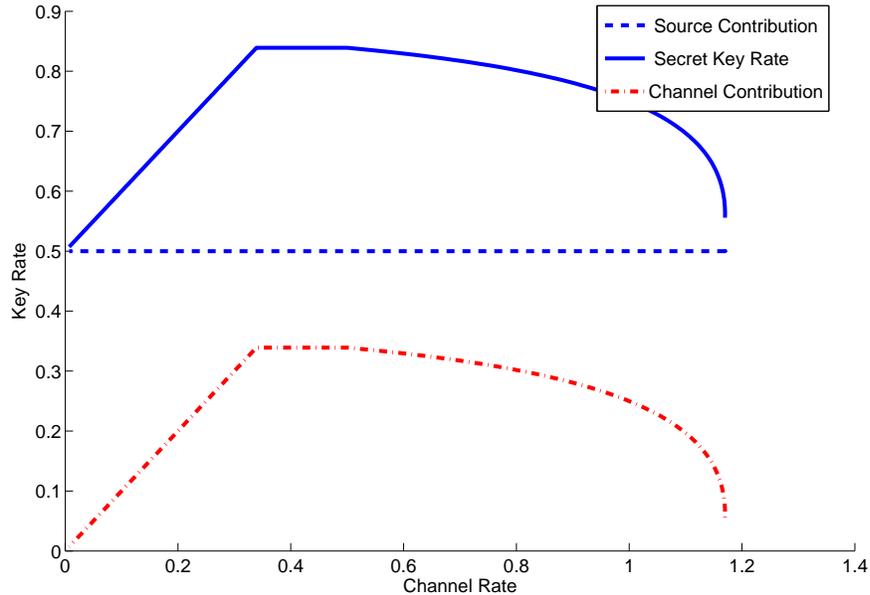}
\caption{Secret-key-rate in the presence of a public discussion channel in the Gaussian example~\eqref{eq:parChan}.  The solid curve is the secret-key-rate, which is the sum of the two other curves. The horizontal line is the key rate from the source components. Regardless of the channel rate, the rate is 0.5 bits/symbol. The dashed-dotted curve is the key-rate  using the channel $I(\rvx;\rvy|\rvz)$. } \label{fig:reqDiscussion}

\vspace{-2em}
\end{figure}

The presence of a public discussion channels allows us to decouple the source and channel codebooks.  We generate two separate keys --- one from the source component using a Slepian-Wolf codebook and one from the channel component using the key-agreement protocol described in~\cite{ahlswedeCsiszar:93,maurer:93}.

The upper bound expression~\eqref{eq:Ckey_PubDisc} in Theorem~\ref{thm:secInteractive} is established using techniques similar to the proof of the upper bound on the secret-key rate for the channel model~\cite[Theorem 3]{ahlswedeCsiszar:93}. A derivation is provided in section~\ref{sec:Discussion}.

Fig.~\ref{fig:reqDiscussion} illustrates the contribution of source and channel coding components for the case of Gaussian parallel channels~\eqref{eq:parChan} consisting of (physically) degraded component channels.  The term $I(\rvu;\rvv)$ is independent of the channel coding rate, and is shown by the horizontal line. The channel equivocation rate $I(\rvx;\rvy|\rvz)$ is maximized at the secrecy capacity. The overall key rate is the sum of the two components. Note that unlike Fig.~\ref{fig:req}, there is no inherent tradeoff between source and channel coding contributions in the presence of public discussion channel and the design of source and channel codebooks is decoupled.

\section{ Achievability: Coding Theorem}
\label{sec:AchievMain} We   demonstrate the coding theorem  in the
special case when $\rva=\rvx$ and $\rvb=0$ in
Lemma~\ref{lem:achievRate}.
Accordingly we have that~\eqref{eq:Rch}
and~\eqref{eq:Req} reduce to
\begin{subequations}
\begin{equation}
R_\mrm{ch} = I(\rvx;\rvy)\label{eq:RchS}
\end{equation}
\begin{equation}
R_\mrm{eq}^- = I(\rvx;\rvy)-I(\rvx;\rvz)\label{eq:ReqS}
\end{equation}
\end{subequations}

The more general case, can be incorporated by introducing an auxiliary channel $\rva\rightarrow\rvx$ and superposition coding~\cite{csiszarKorner} as outlined in Appendix~\ref{app:GenAchiev}. Furthermore, in our discussion below we will assume that the  distributions $p_{\rvt|\rvu}$ and $p_{\rvx}$ are selected such that, for  a sufficiently small but fixed $\del > 0$, we have
\begin{equation}
\label{eq:tightRateCons}
\beta R_\mrm{wz} = R_\mrm{ch} - 3\del.
\end{equation}
We note that the optimization over the joint distributions in Lemma~\ref{lem:achievRate} is over the region $\beta R_\mrm{wz} \le R_\mrm{ch}$. If  the joint distributions satisfy  that $\beta R_\mrm{wz} = \al(R_\mrm{ch}-3\del)$ for some $\al < 1$, one can use the code construction below for a bock-length $\al n$ and then transmit an independent message at rate $R_\mrm{eq}^-$  using a perfect-secrecy wiretap-code. This provides a rate of

$$\al\left(\frac{\beta}{\al}R_\mrm{wz} + R_\mrm{eq}^-\right) + (1-\al)R_\mrm{eq}^- = R_\mrm{eq}^- + \beta R_\mrm{wz},$$
as required.

\subsection{Codebook Construction}
\label{subsec:codebookCons}

\begin{figure}
\psfrag{X}{$\rvx^n$}\psfrag{Y}{$\rvy^n$}
\psfrag{U}{$\rvu^N$}\psfrag{V}{$\rvv^N$}
\psfrag{K}{$\rvk$}
\psfrag{1}{$2^{N(I(\rvt;\rvu)-I(\rvt;\rvv))}$ bins}
\psfrag{2}{$2^{N(\rvt;\rvv)}$ cws/bin}
\includegraphics[scale=0.5]{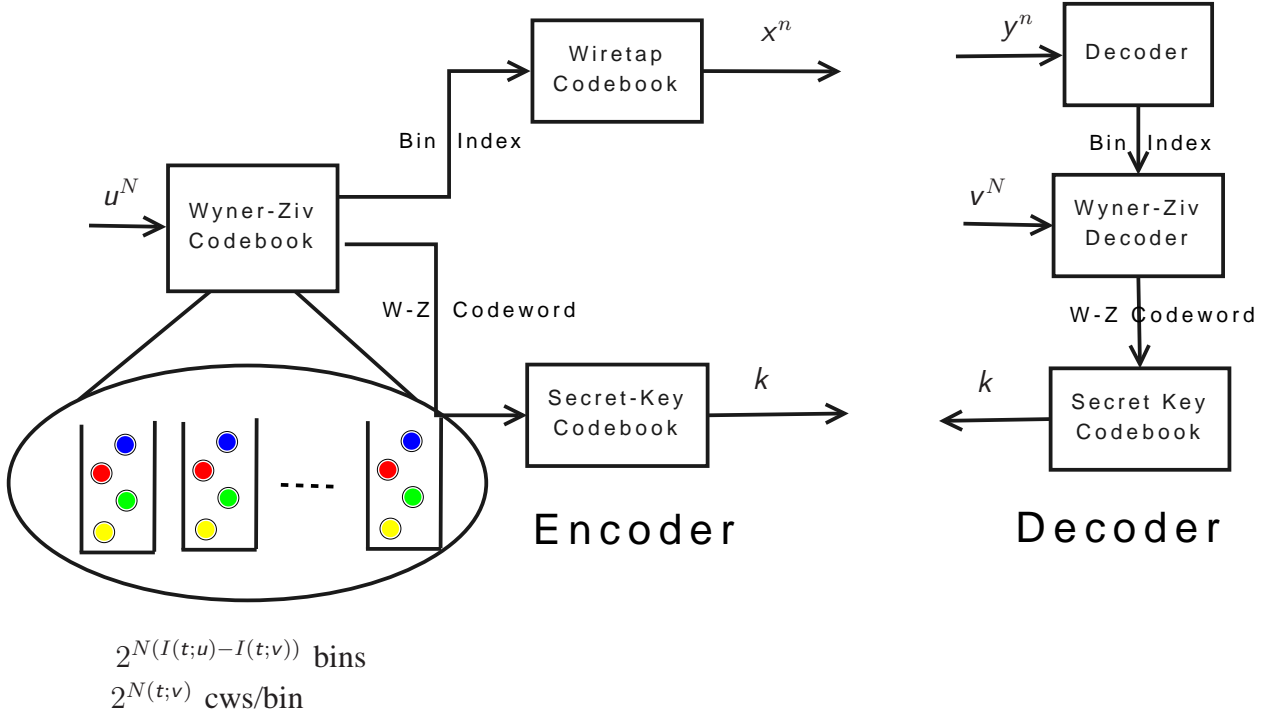}
\caption{Source-Channel Code Design for secret-key distillation problem. The source sequence $\rvu^N$ is mapped to a codeword in a Wyner-Ziv codebook. This codeword determines the secret-key via the secret-key codebook. The bin index of the codeword constitutes a message in the wiretap codebook. }
\label{fig:encFig}
\end{figure}

Our codebook construction is as shown in the Fig.~\ref{fig:encFig}.

An intuition behind the codebook construction is first described.
The wiretap channel carries an ambiguity of $2^{n\{I(\rva;\rvy|\rvb)-I(\rva;\rvz|\rvb)\}}$ at the eavesdropper for each transmitted message. Furthermore, each message only reveals the bin index. Hence
it carries an additional ambiguity of $2^{NI(\rvv;\rvt)}$ codeword sequences. Combining these two effects the total  ambiguity is $2^{n\{I(\rva;\rvy|\rvb)-I(\rva;\rvz|\rvb) + \beta I(\rvv;\rvt)\}}$. Thus a secret-key can be produced at the rate $I(\rva;\rvy|\rvb)-I(\rva;\rvz|\rvb) + \beta I(\rvv;\rvt)$. This heuristic intuition is made precise below.

\begin{figure}
\psfrag{3}{List Size: $2^{n(I(\rvby;\rva|\rvb)-I(\rvz;\rva|\rvb)}$}
\psfrag{4}{$2^{NI(\rvt;\rvv)}$ codewords per bin}
\includegraphics[scale=0.5]{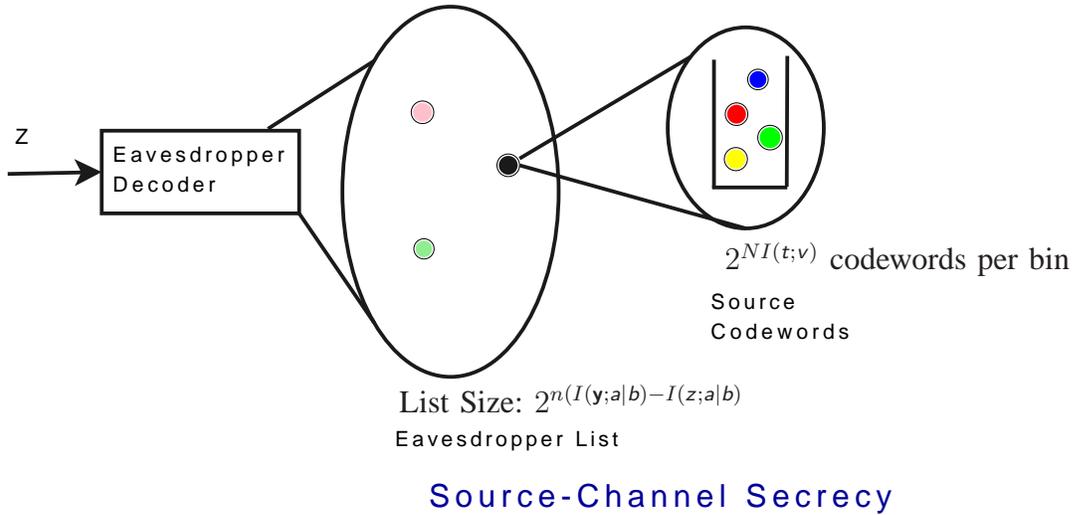}
\label{fig:eaves}
\caption{Equivocation at the eavesdropper through the source-channel codebook.
The channel codebook induces an ambiguity of $2^{n(I(\rva;\rvy|\rvb)- I(\rva;\rvz|\rvb))}$ among the codeword sequences $\rva^n$ when the decoder observes $\rvz^n$. Each sequence $\rva^n$ only reveals the bin index of the Wyner-Ziv codeword. In induces an ambiguity of $2^{N I(\rvt;\rvv)}$ at the eavesdropper, resulting in a total ambiguity of $2^{n(\beta I(\rvt;\rvv)+ I(\rva;\rvy|\rvb))-I(\rva;\rvz|\rvb)}$.}
\end{figure}

The coding scheme consists of three codebooks: Wyner-Ziv codebook, secret-key codebook and a wiretap codebook that are constructed via a random coding construction. In our discussion below we will be using the notion of strong typicality. Given a random variable $\rvt$, the set of all sequences of length $N$ and type that coincides with the distribution $p_\rvt$ is denoted by $T_\rvt^N$. The set of all sequences whose empirical type is in an $\eps$-shell of $p_\rvt$ is denoted by $T_{\rvt,\eps}^N$. The set of jointly typical sequences are defined in an analogous manner. Given a sequence $u^N$ of type $T_\rvu^N$, the set of all sequences $v^N$ that have a joint type of $p_{\rvu,\rvv}()$ is denoted by $T^N_{\rvu,\rvv}(u^N)$.
We will be using the following properties of typical sequences
\begin{subequations}
\begin{align}
|T_{\rvt,\eps}^N| &= \exp(N(H(\rvt)+o_\eps(1))) \label{eq:typeSize}\\
\Pr(\rvt^N = t^N) &= \exp(-N(H(\rvt) + o_\eps(1))), \hspace{1em} \forall~t^N \in T_{\rvt,\eps}^N \label{eq:typeProb}\\
\Pr(\rvt^N \in T_{\rvt,\eps}^N) &\ge 1- o_\eps(1), \label{eq:typeSetProb}
\end{align}
\end{subequations}
where $o_\eps(1)$ is a term that approaches zero as $N\rightarrow \infty$ and $\eps \rightarrow 0$.

For fixed, but sufficiently small constants  $\del > 0$ and $\eta =  \del/\beta > 0$, let,
\begin{subequations}
\begin{align}
M_\mrm{WZ} &= \exp(N (R_\mrm{s} - \eta)) \label{eq:MWZ}\\
N_\mrm{WZ} &= \exp({N(R_\mrm{wz} + 2\eta)}) \label{eq:NWZ}\\
M_\mrm{SK} &= \exp({n(I(\rvx;\rvz)-\del)}) \label{eq:MSK}\\
N_\mrm{SK} &= \exp({n(\beta R_\mrm{s} + R_\mrm{eq}^--\del)})\label{eq:NSK}
\end{align}
\end{subequations}
Substituting~\eqref{eq:Rch}-\eqref{eq:Rwz} and~\eqref{eq:tightRateCons} into \eqref{eq:MWZ}-\eqref{eq:NSK} we have that
\begin{equation}
N_\mrm{tot} \defeq M_\mrm{SK}\cdot N_\mrm{SK} = M_\mrm{WZ} \cdot N_\mrm{WZ} = \exp({N(I(\rvt;\rvu)+\eta)})\label{eq:Tdefn}
\end{equation}

We construct the Wyner-Ziv and secret-key codebooks as follows. Randomly and independently select  $N_\mrm{tot}$ sequences from the set of $\rvt-$typical sequences $T_\rvt^N$. Denote this set $\cT$. Randomly and independently partition this set into the following codebooks\footnote{As will be apparent in the analysis, the only pairwise independence is required between the codebooks i.e., $\forall \rvt^N, \rvht^N \in \cT$, $\Pr\left(\Phi_\mrm{WZ}(\rvt^N) = \Phi_\mrm{WZ}(\rvht^N) |  \Phi_\mrm{SK}(\rvt^N) = \Phi_\mrm{SK}(\rvht^N)\right) = \Pr\left(\Phi_\mrm{WZ}(\rvt^N) = \Phi_\mrm{WZ}(\rvht^N)\right) = \frac{1}{N_\mrm{WZ}}$}:

\begin{itemize}
\item  \emph{Wyner-Ziv} codebook with $N_\mrm{WZ}$ bins consisting of $M_\mrm{WZ}$ sequences. The $j^\mrm{th}$ sequence in bin $i$ is denoted by $\rvt_{ij,\mrm{WZ}}^{N}$.
\item \emph{Secret-key} codebook with $N_\mrm{SK}$ bins consisting of $M_\mrm{SK}$ sequences. The $j^\mrm{th}$ sequence in bin $i$ is denoted by $\rvt_{ij,\mrm{SK}}^{N}$.
\end{itemize}

We define two functions $\Phi_\mrm{WZ}: \cT \rightarrow \{1,\ldots, N_\mrm{WZ}\}$
and  $\Phi_\mrm{SK}: \cT \rightarrow \{1,\ldots, N_\mrm{SK}\}$ as follows.
\begin{defn}
\label{def:encoding_funcs}
Given a codeword sequence $t^N$, define two mappings
\begin{enumerate}
\item $\Phi_\mrm{WZ}(t^N)=i$,
if $\exists j \in [1,M_\mrm{WZ}]$, such that $t^N = \rvt^{N}_{ij,\mrm{WZ}}$.
\item  $\Phi_\mrm{SK}(t^N)=i$, if $\exists j \in [1,M_\mrm{SK}]$ such that
$t^N = \rvt^{N}_{ij,\mrm{SK}}$.
\end{enumerate}
\end{defn}
\hfill$\blacksquare$

The channel codebook consists of $N_\mrm{WZ} = \exp({n(R_\mrm{ch}-\del)})$ sequences $\rvx^n$ uniformly and independently selected from the set of $\rvx-$typical sequences $T_\rvx^n$. The channel encoding function maps message $i$ into the sequence $\rvx_i^n$, i.e., $\Phi_\mrm{ch}: \{1,\ldots, N_\mrm{WZ}\}\rightarrow \cX^n$ is defined as $\Phi_\mrm{ch}(i) = \rvx_i^n$.

\subsection{Encoding}
\label{subsec:Encoding}
Given a source sequence $u^N$, the encoder produces a secret-key $\rvk$ and a transmit sequence $x^N$ as shown in Fig.~\ref{fig:encFig}.
\begin{itemize}
\item Find a sequence $t^N \in \cT$ such that $(u^N,t^N) \in T_{\rvu\rvt,\eps}^N$. Let $\cE_1$ be the even that no such $t^N$ exists.
\item  Compute $\phi = \Phi_\mrm{WZ}(t^N)$ and $\rvk = \Phi_\mrm{SK}(t^N)$.  Declare $\rvk$ as the secret-key.
\item  Compute $x_i^n = \Phi_\mrm{ch}(\phi)$, and transmit this sequence over $n-$uses of the DMC.
\end{itemize}

\subsection{Decoding}
\label{subsec:Decoding} The main steps of decoding at the legitimate
receiver are shown in Fig.~\ref{fig:encFig} and described below.
\begin{itemize}
\item Given a received sequence $y^n$, the sender looks for a
unique index $i$ such that $(x_i^n,y^n) \in T^n_{\rvx\rvy,\eps}$. An
error event $\cE_2$ happens if $x_i^n$ is not  the transmitted
codeword.

\item Given the observed source sequence $v^N$, the decoder then searches
for a unique index $j \in [1,M_\mrm{WZ}]$ such that
$(t_{ij,\mrm{WZ}}^N, v^N) \in T_{\rvt\rvv,\eps}^N$. An error event
$\cE_3$ is declared if a unique index does not exist.

\item The decoder computes $\rvhk = \Phi_\mrm{SK}(t_{ij,\mrm{WZ}}^N)$ and declares $\rvhk$ as the secret key.
\end{itemize}

\subsection{Error Probability Analysis}
\label{subsec:ErrorProb}
The error event of interest is $\cE = \{\rvk \neq \rvhk\}$. We argue  that   selecting $n \rightarrow \infty$ leads to $\Pr(\cE)\rightarrow 0$.

In particular, note that $\Pr(\cE) = \Pr(\cE_1 \cup \cE_2 \cup
\cE_3) \le \Pr(\cE_1) + \Pr(\cE_2) + \Pr(\cE_3)$. We argue that each
of the terms vanishes with $n \rightarrow \infty$.

Recall that $\cE_1$ is the event that the encoder does not find a
sequence in $\cT$ typical with $\rvu^N$. Since $\cT$ has
$\exp({N(I(\rvu;\rvt)+\eta)})$ sequences randomly and uniformly
selected from the set $T_{\rvt}^N$, we have that
$\Pr(\cE_1)\rightarrow 0$.

Since the number of channel codewords equals $N_\mrm{WZ} =
\exp({n(I(\rvx;\rvy)-\del)})$, and the codewords are selected
uniformly at random from the set $T_{\rvx,\eps}^n$, the error event
$\Pr(\cE_2) \rightarrow 0$.

Finally, since the number of sequences in each bin satisfies $M_\mrm{WZ} = \exp({N(I(\rvt;\rvv)-\eta)})$, joint typical decoding guarantees that $\Pr(\cE_3) \rightarrow 0$.

\subsection{Secrecy Analysis}
\label{subsec:SecAnalysis}
In this section, that for the coding scheme discussed above, the equivocation at the eavesdropper is close (in an asymptotic sense) to $R_\mrm{key}$.

First we establish some uniformity properties which will be used in the subsequent analysis.

\subsubsection{Uniformity  Properties}

In our code construction $\Phi_\mrm{WZ}$  satisfies some useful properties which will be used in the sequel.

\begin{lemma}
The random variable $\Phi_\mrm{WZ}$ in Def.~\ref{def:encoding_funcs} satisfies the following relations
\begin{subequations}
\begin{align}
&\frac{1}{n}H(\Phi_\mrm{WZ}) = \beta R_\mrm{WZ} + o_{\eta} (1)\label{eq:PhiWZUnif}\\
&\frac{1}{n}H(\rvt^N|\Phi_\mrm{WZ}) = \beta I(\rvt;\rvv) + o_{\eta} (1)\label{eq:PhitWZUnif}\\
&\frac{1}{n}H(\Phi_\mrm{WZ}| \rvz^n) = I(\rvx;\rvy) - I(\rvx;\rvz) + o_\eta(1)\label{eq:PhiWiretapEquiv}
\end{align}\end{subequations}
\label{lem:Uniformity}
\end{lemma}
where $o_\eta(1)$ vanishes to zero as we take $\eta \rightarrow 0$ and $N\rightarrow \infty$ for each $\eta$.

\begin{new-proof}
Relations~\eqref{eq:PhiWZUnif} and~\eqref{eq:PhitWZUnif} are established below by using the properties of typical sequences (c.f.~\eqref{eq:typeSize}-\eqref{eq:typeSetProb}).  Relation~\eqref{eq:PhiWiretapEquiv} follows from the secrecy analysis of the channel codebook when the message is $\Phi_\mrm{WZ}$. The details can be found in e.g.,~\cite{wyner:75Wiretap}.

To establish~\eqref{eq:PhiWZUnif}, define the function $\G_\mrm{WZ}: \cT \rightarrow \{1,\ldots,M_\mrm{WZ} \}$ to identify the position of the sequence $\rvt^N \in \cT$ in a given bin i.e., $\G_\mrm{WZ}(\rvt^N_{ij,\mrm{WZ}})= j$ and note that,
\begin{align}
& \Pr(\G_\mrm{WZ}=j,\Phi_\mrm{WZ}=i) \le
\sum_{\rvu^N \in T_{\rvu,\rvt,\eta}(t_{ij,\mrm{WZ}}^N)}\Pr(\rvu^N)\label{eq:indepStep1}\\
&=  \sum_{\rvu^N \in T_{\rvu,\rvt,\eta}(t_{ij,\mrm{WZ}}^N)}\exp({-N(H(\rvu)+ o_\eta(1))}) \label{eq:indepStep2}\\
&= \exp({N(H(\rvu|\rvt)+ o_\eta(1))})\exp({-N(H(\rvu)+ o_\eta(1))}) \label{eq:indepStep3}\\
&= \exp({-N(I(\rvt;\rvu) + o_\eta(1))})\label{eq:indepRes}
\end{align}
where~\eqref{eq:indepStep1} follows from the construction of the joint-typicality encoder,~\eqref{eq:indepStep2} from~\eqref{eq:typeProb} and~\eqref{eq:indepStep3} from~\eqref{eq:typeSize}.
Marginalizing~\eqref{eq:indepStep1}, we have that
\begin{align}
\Pr(\Phi_\mrm{WZ}=i)&= \sum_{j=1}^{M_\mrm{WZ}}\Pr(\G_\mrm{WZ}=j,\Phi_\mrm{WZ}=i) \notag\\
&\le M_\mrm{WZ}\exp({-N(I(\rvt;\rvu) + o_\eta(1))}) \notag\\
&= \exp(-N(I(\rvt;\rvu)-I(\rvt;\rvv)+o_\eta(1))) \notag\\
&= \exp(-N(R_\mrm{WZ} + o_\eta(1)))\label{eq:margPhiWZ}
\end{align}

Eq.~\eqref{eq:PhiWZUnif} follows from~\eqref{eq:margPhiWZ} and the continuity of the entropy function.  Furthermore, we have from~\eqref{eq:indepRes} that
\begin{equation}\label{eq:jointDis}\frac{1}{N}H(\Phi_\mrm{WZ},\G_\mrm{WZ})= I(\rvt;\rvu)+o_\eta(1).\end{equation} The relation~\eqref{eq:PhitWZUnif} follows
by substituting~\eqref{eq:PhiWZUnif}, since

\begin{multline}
\frac{1}{N}H(\rvt^N |\Phi_\mrm{WZ}) = \frac{1}{N}H(\G_\mrm{WZ} |\Phi_\mrm{WZ}) =
 \frac{1}{N}H(\G_\mrm{WZ},\Phi_\mrm{WZ}) - \frac{1}{N}H(\Phi_\mrm{WZ}) = I(\rvt;\rvv) + o_\eta(1).
\end{multline}

\end{new-proof}

\begin{lemma}
The construction of the secret-key codebook and Wyner-Ziv codebook is such that the eavesdropper can decode the sequence $\rvt^N$ if it is revealed the secret-key $\Phi_\mrm{SK}=\rvk$ in addition to its observed sequence $\rvz^n$. In particular

\begin{equation}\frac{1}{n}H(\rvt^N|\rvz^n,\rvk) = o_\eta(1).\end{equation}
\label{lem:FanoLem}
\end{lemma}

\begin{new-proof}
We show that there exists a decoding function $g: \cZ^n \times \{1,2,\ldots, N_\mrm{SK}\} \rightarrow \cT$ that such that $\Pr(\rvt^N \neq g(\rvz^n,\rvk)) \rightarrow 0$ as $n\rightarrow \infty$. In particular, the decoding function $g(\cdot,\cdot)$ searches for the sequences in the bin associated with $\rvk$ in the secret-key codebook, whose bin-index in the Wyner-Ziv codebook maps to a sequence $\rvx_i^n$ jointly typical with the received sequence $\rvz^n$. More formally, \begin{itemize}
\item Given $\rvz^n$, the decoder constructs a the set of indices $\cI_\mrm{x} = \{i: (\rvx_i^n,\rvz^n) \in T_{\rvx\rvz,\eps}^n\} $.
\item  Given $\rvk$, the decoder constructs a set of sequences,  $\cS =\left\{\rvt_{\rvk j,\mrm{SK}}^N:    \Phi_\mrm{WZ}(\rvt^N_{\rvk j,\mrm{SK}}) \in \cI_\mrm{x}, 1\le j \le M_\mrm{SK},\right\}$.
\item   If $\rvS$ contains a unique sequence $\rvht^N$, it is declared to be the required sequence. An error event is defined as
\begin{align}
\cJ &= \{\rvht^N \neq \rvt^N\}\notag\\
&= \left\{\exists j, 1\le j \le M_\mrm{SK}, \Phi_\mrm{WZ}(\rvt^N_{\rvk,j,\mrm{SK}})\in \cI_\mrm{x}, j \neq j_0 \right\},\label{eq:EavesErrEvent}
\end{align}
where $j_0$ is the index of the sequence $\rvt^N$ in bin $\rvk$ of the secret-key codebook, i.e.,  $\rvt^N_{\rvk j_0,\mrm{SK}}=\rvt^N$.
\end{itemize}

It suffices to show that $\Pr(\cJ) \rightarrow 0 $ as $n \rightarrow \infty$.

We begin by defining the following events:

\begin{itemize}

\item  The event that the sequence $\rvt^N \notin \cS$, which is equivalent to
$$\cJ_0 = \left\{\Phi_\mrm{WZ}(\rvt^N_{\rvk,j_0,\mrm{SK}})\notin \cI_\mrm{x} \right\}.$$
From~\eqref{eq:typeSetProb} we have that
$\Pr(\cJ_0) = o_\eta(1)$.

\item For each $j=1,2,\ldots M_\mrm{SK}, j\neq j_0$
the event $\cJ_j$  that the sequence $\rvt_{\rvk j\mrm{SK}}^N \in \cS$, $$\cJ_j = \left\{\Phi_\mrm{WZ}(\rvt^N_{\rvk,j,\mrm{SK}})\in \cI_\mrm{x} \right\}.$$

\item  For each $j=1,2,\ldots M_\mrm{SK}, j\neq j_0$,
define the collision event  that $\rvt_\mrm{\rvk j,\mrm{SK}}^N$ and
$\rvt_\mrm{\rvk j_0,\mrm{SK}}^N$ belong to the same bins in the in the Wyner-Ziv codebook  $$\cJ_\mrm{col,j} = \left\{\Phi_\mrm{WZ}(\rvt^N_{\rvk j,\mrm{SK}}) = \Phi_\mrm{WZ}(\rvt^N_{\rvk j_0,\mrm{SK}}) \right\}.$$
\end{itemize}

Now we upper bound the error probability in terms of these events.
\begin{align}
\Pr(\cJ) &\le \Pr(\cJ |\cJ_0^c) + \Pr(\cJ_0)\notag \\
&\le \sum_{j=1, j \neq j_0 }^{M_\mrm{SK}}\Pr(\cJ_j |\cJ_0^c) + o_\eta(1),\label{eq:UnionBoundEvent}
\end{align}

Now observe that
\begin{align}
\Pr(\cJ_j | \cJ_0^c) &= \Pr(\cJ_j  \cap \cJ_\mrm{col,j}^c| \cJ_0^c)
+ \Pr(\cJ_j  \cap \cJ_\mrm{col,j}| \cJ_0^c)\\
&\le \Pr(\cJ_j  \cap \cJ_\mrm{col,j}^c| \cJ_0^c) + \Pr(\cJ_\mrm{col,j}| \cJ_0^c) \notag\\
&\le \Pr(\cJ_j | \cJ_0^c \cap \cJ_\mrm{col,j}^c) + \Pr(\cJ_\mrm{col,j}|\cJ_0^c).
\label{eq:Bound2}
\end{align}
We bound each of the two terms in~\eqref{eq:Bound2}.
The first term is conditioned on the event that the sequences $\rvt_{\rvk j,\mrm{SK}}^N$ and $\rvt_{\rvk j_0,\mrm{SK}}^N$  are assigned to independent bins in the Wyner-Ziv codebook.  This event is equivalent to the event that a randomly selected sequence $\rvx^N$  belongs to the typical set $\cI_\mrm{x}$. The error event is bounded as~\cite{coverThomas}

\begin{equation}
\label{eq:NoCollError}
\Pr(\cJ_j | \cJ_0^c \cap \cJ_\mrm{col,j}^c) \le \exp(-n(I(\rvx;\rvz)-3\eps)).
\end{equation}

To upper bound the second term,
\begin{align}
&\Pr(\cJ_j | \cJ_0^c) = \Pr(\cJ_j) \label{eq:indepEvnts} \\
&= \exp(-n (\beta R_\mrm{WZ} + 2\delta)) \label{eq:WZBin} \\
&= \exp(-n (I(\rvx;\rvy) - \delta)) \label{eq:CollisionError}
\end{align}
where~\eqref{eq:indepEvnts} follows from the fact the event $\cJ_0$
is due to the atypical channel behavior and is independent of the
random partitioning event that induces $\cJ_j$,~\eqref{eq:WZBin}
follows from the fact that each sequence is independently assigned
to one of $\exp\{n(\beta R_\mrm{WZ} + 2\del)\}$ bins in the code
construction and~\eqref{eq:CollisionError} follows via
relation~\eqref{eq:tightRateCons}.

Substituting~\eqref{eq:CollisionError} and~\eqref{eq:NoCollError} into~\eqref{eq:Bound2}, we have

\begin{align}
\Pr(\cJ_j | \cJ_0^c) &
\le \exp(-n(I(\rvx;\rvz)-3\eps))+ \exp(-n ( I(\rvx;\rvy) - \delta))   \notag \\
& \le \exp(-n(I(\rvx;\rvz)-4\eps)), \quad n \ge n_0 \label{eq:TermBound},
\end{align}
where we use the fact that $I(\rvx;\rvy)> I(\rvx;\rvz)$ in the last step so that the required $n_0$ exists.

Finally substituting~\eqref{eq:TermBound} into~\eqref{eq:UnionBoundEvent} and using relation~\eqref{eq:MSK} for $M_\mrm{SK}$, we have that
\begin{align}
\Pr(\cJ) \le \exp(-n(\del - 4\eps)) + o_\eta(1),
\end{align}
which vanishes with $n$, whenever the decoding function selects $\eps < \del/4$.
\end{new-proof}

\subsubsection{Equivocation Analysis}
\label{subsubsec:EquivAnalysis}
It remains to show that the equivocation rate at the eavesdropper approaches the secret-key rate as $n \rightarrow \infty$, which we do below.

\begin{align}
H(\rvk | \rvz^n) &= H(\rvk, \rvt^N | \rvz^n) - H(\rvt^N |
\rvz^n, \rvk) \notag\\
&= H(\rvt^N | \rvz^n) - H(\rvt^N | \rvz^n,\rvk) \label{eq:VGdetFunT}\\
&= H(\rvt^N,\Phi_\mrm{WZ} | \rvz^n) - H(\rvt^N | \rvz^n, \rvk)
\label{eq:VFdetFunT}\\
&=  H(\rvt^N | \Phi_\mrm{WZ}, \rvz^n)  + H(\Phi_\mrm{WZ}| \rvz^n) -
H(\rvt^N | \rvz^n, \rvk) \notag\\
&=  H(\rvt^N | \Phi_\mrm{WZ})  + H(\Phi_\mrm{WZ}| \rvz^n) - H(\rvt^N |
\rvz^n, \rvk), \label{eq:VTPhMarkov}\\
&= n\beta I(\rvt;\rvv) + n \{I(\rvx;\rvy)-I(\rvx;\rvz)\} + n o_\eta(1)\label{eq:Equiv} \\
&= n(R_\mrm{key} + o_\eta(1)),
\end{align}
where~\eqref{eq:VGdetFunT} and~\eqref{eq:VFdetFunT} follow from the
fact that $\Phi_\mrm{WZ}$ is a deterministic
function of $\rvt^N$ and~\eqref{eq:VTPhMarkov} follows from the fact
that $\rvt^N \rightarrow \Phi_\mrm{WZ} \rightarrow \rvz^n$ holds for
our code construction. and~\eqref{eq:Equiv} step follows from ~\eqref{eq:PhitWZUnif} and~\eqref{eq:PhiWiretapEquiv} in Lemma~\ref{lem:Uniformity} and
 Lemma~\ref{lem:FanoLem}.

\section{Proof of the Upper bound (Lemma~\ref{lem:upperBound})}
\label{sec:UpperBound}

Given a sequence of $(n,N)$ codes that achieve a secret-key-rate
$R_\mrm{key}$, there exists a sequence $\eps_n$, such that
$\eps_n\rightarrow 0$ as $n\rightarrow \infty$, and
\begin{subequations}\begin{align}
&\frac{1}{n}H(\rvk | \rvy^n, \rvv^N) \le \eps_n \label{eq:FanoLemma}\\
&\frac{1}{n}H(\rvk | \rvz^n ) \ge \frac{1}{n}H(\rvk) - \eps_n.
\label{eq:EquivCond}
\end{align}
\end{subequations}
We can now upper bound the rate $R_\mrm{key}$ as follows.
\begin{align}
nR_\mrm{key} &= H(\rvk) \notag\\
&= H(\rvk|\rvy^n, \rvv^N) + I(\rvk;\rvy^n,\rvv^N)\notag\\
&\le n\eps_n  + I(\rvk;\rvy^n,\rvv^N) - I(\rvk; \rvz^n) + I(\rvk;\rvz^n)\label{eq:fanostep}\\
&\le 2n\eps_n  + I(\rvk;\rvy^n,\rvv^N) - I(\rvk; \rvz^n)\label{eq:eqivstep}\\
&= 2n\eps_n + I(\rvk;\rvy^n)-I(\rvk;\rvz^n) + I(\rvk;\rvv^N|\rvy^n)\notag\\
&\le 2n\eps_n + I(\rvk;\rvy^n)-I(\rvk;\rvz^n) + I(\rvk,\rvy^n;\rvv^N)
\label{eq:diff1}
\end{align}
where~\eqref{eq:fanostep} and~\eqref{eq:eqivstep} follow
from~\eqref{eq:FanoLemma} and~\eqref{eq:EquivCond} respectively.

Now, let $J$ be a random variable uniformly distributed over the set
$\{1,2,\ldots, N\}$ and independent of everything else. Let $\rvt_i
= (\rvk,\rvy^n,\rvv_{i+1}^{N},\rvu_1^{i-1})$ and $\rvt =
(\rvk,\rvy^n,\rvv_{J+1}^{N},\rvu_1^{J-1},J)$, and $\rvv_J$ be a
random variable that conditioned on $J=i$ has the distribution of
$p_{v_i}$. Note that since $\rvv^N$ is memoryless,  $\rvv_J$ is
independent of $J$ and has the same marginal distribution as $\rvv$. Also note that $\rvt \rightarrow \rvu_\mrm{J} \rightarrow \rvv_\mrm{J}$ holds.

\begin{align}
I(\rvk,\rvy^n;\rvv^N) &= \sum_{i=1}^n I(\rvk, \rvy^n; \rvv_i |
\rvv_{i+1}^n)\notag\\
&\le \sum_{i=1}^N I(\rvk, \rvy^n, \rvv_{i+1}^{n}; \rvv_i  )\notag\\
&\le \sum_{i=1}^N I(\rvk, \rvy^n, \rvv_{i+1}^{n},\rvu_{1}^{i-1}; \rvv_i  )\notag\\
&= NI(\rvk, \rvy^n, \rvv_{J+1}^{n},\rvu_{1}^{J-1}; \rvv_J|J
)\notag\\
&= N I(\rvk, \rvy^n, \rvv_{J+1}^{n},\rvu_{1}^{J-1},J;\rvv_J)  -
I(J;\rvv_J)\notag \\&=N I(\rvt; \rvv  )\label{eq:vUB}
\end{align}
where~\eqref{eq:vUB} follows from the fact that $\rvv_J$ is
independent of $J$ and has the same marginal distribution as $\rvv$.

Next, we upper bound $I(\rvk;\rvy^n)-I(\rvk;\rvz^n)$ as below. Let
$p_{\rvx_i}$ denote the channel input distribution at time $i$ and
let $p_{\rvy_i,\rvz_i}$ denote the corresponding output
distribution. Let $p_\rvx = \frac{1}{n}\sum_{i=1}^n p_{\rvx_i}$ and
let $p_\rvy$ and $p_\rvz$ be defined similarly.
\begin{align}
I(\rvk;\rvy^n)-I(\rvk;\rvz^n) &\le I(\rvk;\rvy^n|\rvz^n)\notag\\
&\le I(\rvx^n;\rvy^n|\rvz^n)\label{eq:markovCh}\\
&\le \sum_{i=1}^n I(\rvx_i;\rvy_i |\rvz_i)\label{eq:chMem}\\
&\le nI(\rvx;\rvy|\rvz),\label{eq:chUB}
\end{align}
where~\eqref{eq:markovCh} follows from the Markov condition
$\rvk\rightarrow \rvx^n \rightarrow (\rvy^n,\rvz^n)$
and~\eqref{eq:chMem} follows from the fact that the channel is
memoryless and~\eqref{eq:chUB} follows from Jensen's inequality since the term $I(\rvx;\rvy|\rvz)$ is concave in the distribution $p_\rvx$ (see e.g.,~\cite[Appendix-I]{khistiTchamWornell:07}).

Combining~\eqref{eq:chUB} and~\eqref{eq:vUB} we have that
\begin{equation}
R_\mrm{key} \le I(\rvx;\rvy|\rvz) + \beta I(\rvv;\rvt),
\end{equation}
thus establishing the first half of the condition in
Lemma~\ref{lem:upperBound}. It remains to show that the condition
$$\beta\{I(\rvt;\rvu)-I(\rvt;\rvv)\}\le I(\rvx;\rvy)$$
is also satisfied. Since $\rvu^N\rightarrow \rvx^n\rightarrow \rvy^n$
holds, we have that
\begin{align}
nI(\rvx;\rvy) &\ge I(\rvx^n;\rvy^n)\\
&\ge I(\rvu^N;\rvy^n)\\
&\ge I(\rvu^N;\rvy^n,\rvk) - I(\rvv^N;\rvy^n,\rvk) -
n\eps_n,\label{eq:CapUB}
\end{align}
where the last inequality holds, since
\begin{align*}
I(\rvu^N;\rvk|\rvy^n) - I(\rvv^N;\rvy^n,\rvk) &= -I(\rvv^N;\rvy^n)+ I(\rvu^N;\rvk|\rvy^n)- I(\rvv^N;\rvk|\rvy^n)\notag\\
&\le I(\rvu^N;\rvk|\rvy^n)- I(\rvv^N;\rvk|\rvy^n)\\
&=  H(\rvk|\rvy^n,\rvv^N) - H(\rvk|\rvy^n,\rvu^N)\\
&\le n\eps_n,
\end{align*}
where the last step holds via~\eqref{eq:FanoLemma} and the fact that
$H(\rvk|\rvy^n,\rvu^N)\ge 0$.

Continuing~\eqref{eq:CapUB}, we have

\begin{align}
n I(\rvx;\rvy)&\ge I(\rvu^N;\rvy^n,\rvk) - I(\rvv^N;\rvy^n,\rvk) -
n\eps_n\label{eq:RateConsStart}\\
&= \sum_{i=1}^N
\{I(\rvu_i;\rvy^n,\rvk,\rvu_1^{i-1}\rvv_{i+1}^n)-I(\rvv_i;\rvy^n,\rvk,\rvu_1^{i-1}\rvv_{i+1}^n) \} + n\eps_n \label{eq:mutInfDiff}\\
&=
N\{I(\rvu_J;\rvy^n,\rvk,\rvu_1^{J-1}\rvv_{J+1}^n|J)-I(\rvv_J;\rvy^n,\rvk,\rvu_1^{J-1}\rvv_{J+1}^n|J)+\eps_n\} \notag\\
&= N\{I(\rvu_J; \rvt) - I(\rvv_J;\rvt) + I(\rvv_J;J)-I(\rvu_J;J) +
\eps_n\}\notag\\
&= N\{I(\rvu; \rvt) - I(\rvv;\rvt) +\eps_n\}\label{eq:RateConsStop}
\end{align}
where~\eqref{eq:mutInfDiff} follows from the well known chain rule for difference between mutual information expressions (see e.g.,~\cite{elGamal:03}), \eqref{eq:RateConsStop} again follows from the fact that the random
variables $\rvv_J$ and $\rvu_J$ are independent of $J$ and have the
same marginal distribution as $\rvv$ and  $\rvu$ respectively.

The cardinality bound on $\rvt$ is obtained via Caratheordory's
theorem and will not be presented here.

Finally, since the upper bound expression does not depend on the
joint distribution of $(\rvt,\rvx)$,  it suffices to optimize over
those distributions where $(\rvt,\rvx)$ are independent.

\section{Reversely Degraded Channels}
\label{sec:ReverselyDegraded}
\subsection{Proof of Theorem~\ref{corol:RevDegCap}}
\label{subsec:ParChanProof}
First we show that the expression is an upper bound on the capacity. From Lemma~\ref{lem:upperBound}, we have that
$$C_\mrm{key} \le \max_{(\rvx, \rvt)} I(\rvx;\rvy|\rvz) + \beta I(\rvt;\rvv),$$
where we maximize over those distributions where $(\rvx,\rvt)$ are
mutually independent, $\rvt\rightarrow\rvu\rightarrow\rvv$, and
$$I(\rvx;\rvy) \ge \beta(I(\rvt;\rvu)-I(\rvt;\rvv)).$$
For the reversely degraded parallel independent channels, note that
\begin{align*}
I(\rvx;\rvy) &\le \sum_{i=1}^M I(\rvx_i;\rvy_i)\\
I(\rvx;\rvy|\rvz) &\le \sum_{i=1}^M I(\rvx_i;\rvy_i|\rvz_i),
\end{align*}
with equality when $(\rvx_1,\ldots,\rvx_M)$ are mutually
independent. Thus it suffices to take $(\rvx_1,\ldots,\rvx_M)$ to be
mutually independent, which establishes that the proposed expression
is an upper bound on the capacity.

For achievability, we propose a choice of auxiliary random
variables $(\rva,\rvb)$ in Lemma~\ref{lem:achievRate}, such that the
resulting expression reduces to the capacity. In particular, assume
without loss in generality that for the first $P$ channels we have
that $\rvx_i \rightarrow \rvy_i \rightarrow \rvz_i$ and for the
remaining channels we have that $\rvx_i \rightarrow \rvz_i
\rightarrow \rvy_i$.  Let $a = (\rvx_1,\rvx_2,\ldots, \rvx_M)$ and
$b = (\rvx_{P+1},\ldots, \rvx_M)$ where the random variables
$\{\rvx_i\}$ are mutually independent.  It follows
from~\eqref{eq:Rch} and~\eqref{eq:Req} that
\begin{align}
R_\mrm{ch} &= \sum_{i=1}^MI(\rvx_i;\rvy_i)\\
R_\mrm{eq}^- &= \sum_{i=1}^P I(\rvx_i;\rvy_i|\rvz_i) = \sum_{i=1}^M
I(\rvx_i;\rvy_i|\rvz_i),
\end{align}
where the last equality follows since for $\rvx_i \rightarrow
\rvz_i\rightarrow \rvy_i$, we have that $I(\rvx_i;\rvy_i|\rvz_i)=0$.
Substituting in~\eqref{eq:RLB} and~\eqref{eq:achievIneq} we recover
the capacity expression.

\subsection{Gaussian Case (Corollary~\ref{corol:RevDegCapGauss})}
\label{subsec:ParChanGaussProof}
For the Gaussian case we show that Gaussian codebooks achieve the capacity as in Corollary~\ref{corol:RevDegCapGauss}.

Recall that the capacity expression involves maximizing over random variables $\rvbx = (\rvx_1,\ldots, \rvx_M)$, and $\rvt\rightarrow \rvu \rightarrow \rvv$,
\begin{align}
C_\mrm{key} = \sum_{i} I(\rvx_i;\rvy_i|\rvz_i) + \beta I(\rvt;\rvv)  \label{eq:Robj}
\end{align}
subjected to the constraint that $E[\sum_{i=1}^M \rvx_i^2] \le P$ and
\begin{align}
\sum_i I(\rvx_i;\rvy_i) \ge \beta\{I(\rvt;\rvu)-I(\rvt;\rvv)\}.\label{eq:Rconst}
\end{align}

Let us first  fix the distribution $p_\rvbx$ and upper bound the objective function~\eqref{eq:Robj}. Let  $R \defeq \frac{1}{\beta}\sum_{i=1}^M I(\rvx_i;\rvy_i)$ and $\rvv=\rvu + \rvs$, where $\rvs \sim \cN(0,S)$ is independent of $\rvu$. We will use the conditional entropy power inequality
\begin{equation}
\exp(2h(\rvu + \rvs | \rvt)) \ge \exp(2h(\rvu|\rvt)) +  \exp(2h(\rvs))\label{eq:EPI}
\end{equation}
for any pair of random variables $(\rvt,\rvu)$  independent of $\rvs$. The equality happens if $(\rvu,\rvt)$ are jointly Gaussian.

Note that we can express~\eqref{eq:Rconst} as
\begin{align}
R + h(\rvv)-h(\rvu) &\ge h(\rvv|\rvt) -h(\rvu|\rvt)\\
&= h(\rvu + \rvs | \rvt) - h(\rvu|\rvt)\\
&\ge \frac{1}{2}\log\left(\exp(2h(\rvu|\rvt)) + 2\pi e S\right) - h(\rvu|\rvt)
\end{align}
Letting \begin{equation}\label{eq:D_def} h(\rvu|\rvt)) = \frac{1}{2}\log 2 \pi e D,\end{equation} we have that

\begin{align}
\label{eq:Dineq}
D \ge \frac{S}{\exp(2(R + h(\rvv)- h(\rvu))) -1 }.
\end{align}

Rearranging we have that
\begin{equation}
\sum_{i=1}^M I(\rvx_i;\rvy_i)  \ge \frac{\beta}{2}\left[\log\left(1+\frac{S}{D}\right)-\log(1+S)\right].
\end{equation}

The term $I(\rvt;\rvv)$ in the objective function~\eqref{eq:Robj} can be upper bounded as
\begin{align}
I(\rvt;\rvv) &= h(\rvv) - h(\rvv|\rvt) \notag\\
&= h(\rvv) - h(\rvu + \rvs |\rvt) \notag\\
&\le h(\rvv) - \frac{1}{2}\log(\exp(2h(\rvu|\rvs)) + 2\pi e S)\label{eq:EPI_a}\\
&= \frac{1}{2}\log \frac{1 + S}{D+S}\label{eq:D_a}
\end{align}
where~\eqref{eq:EPI_a} follows by the application of the EPI~\eqref{eq:EPI} and~\eqref{eq:D_a} follows via~\eqref{eq:D_def}.
Thus the objective function~\eqref{eq:Robj} can be expressed as
\begin{align}
C_\mrm{key} = \sum_i I(\rvx_i;\rvy_i|\rvz_i) + \frac{\beta}{2}\log \frac{1 + S}{D+S},
\end{align}
where $D$ satisfies \eqref{eq:Dineq}.

It remains to show that the optimal $\rvbx$ has a Gaussian distribution. Note that the set of feasible distributions for $\rvbx$ is closed and bounded and hence an optimum exists. Also if $p_\rvbx$ is any optimum distribution, we can increase both $R$ and $I(\rvx_i;\rvy_i|\rvz_i)$ by replacing $p_\rvbx$ with a Gaussian distribution (see e.g.,~\cite{khistiWornell:07}) with the same second order moment. Since the objective function is increasing in both these terms, it follows that a Gaussian $p_\rvbx$ also maximizes the objective function~\eqref{eq:Robj}.

\section{Side information at the Wiretapper}
\label{sec:WSI}
We now provide an achievability and a converse for the capacity
stated in Theorem~\ref{lem:WSI}

\subsection{Achievability}

Our coding scheme is a natural extension of the case when $\rvw = 0$.

Since we are only considering  degraded channels note that $R_\mrm{ch}$ and $R_\mrm{eq}^-$ in~\eqref{eq:Rch} and~\eqref{eq:Req} are defined as \begin{align}
R_\mrm{ch} &= I(\rvx;\rvy)\\
R_\mrm{eq}^- &= I(\rvx;\rvy) - I(\rvx;\rvz) = I(\rvx;\rvy|\rvz).
\end{align}
Furthermore, we replace $R_\mrm{s}$ in~\eqref{eq:Rs} with
\begin{equation}
R_\mrm{s} = I(\rvt;\rvv) - I(\rvt;\rvw)\label{eq:RsWSI}
\end{equation}and the secret-key rate in~\eqref{eq:RLB} is
\begin{equation}
R_\mrm{LB} = \beta\{I(\rvt;\rvv)-I(\rvt;\rvw)\}+ I(\rvx;\rvy|\rvz).
\label{eq:RLBWSI}
\end{equation}

The  construction of Wyner-Ziv codebook and wiretap codebook in
Fig.~\ref{fig:encFig} is as discussed in
section~\ref{subsec:codebookCons},~\ref{subsec:Encoding},and~\ref{subsec:Decoding}.
The Wyner-Ziv codebook consists of $\approx 2^{NI(\rvt;\rvu)}$
codeword sequences sampled uniformly from the set $T_t^N$. These
sequences are uniformly and randomly partitioned into $\approx
2^{N\{I(\rvt;\rvu)-I(\rvt;\rvv)\}}$ bins so that there are $\approx
2^{NI(\rvt;\rvv)}$ sequences in each bin. The bin index of a
codeword sequence, $\Phi_\mrm{WZ}$, forms a message for the wiretap
codebook as before.   The construction of the secret key codebook is
modified to reflect the side information sequence at the
eavesdropper.  In particular we construct the secret-key codebook
with parameters
\begin{align}
M_\mrm{SK} &= \exp\left(n(I(\rvx;\rvz)+ \beta I(\rvw;\rvt)) - \del\right)\\
N_\mrm{SK} &= \exp\left(n(\beta R_\mrm{s} + R_\mrm{eq}^- - \del)\right)
\end{align}
and $R_\mrm{s}$ is defined in~\eqref{eq:RsWSI}.

\subsection{Secrecy Analysis}
We show that the equivocation condition at the eavesdropper~\eqref{eq:EquivWSI} holds for the code construction. This is equivalent to showing that
\begin{align}
\frac{1}{n}H(\rvk | \rvw^N,\rvz^n) =
\beta(I(\rvt;\rvv)-I(\rvt;\rvw)) + I(\rvx;\rvy|\rvz)+ o_\eta(n), \label{eq:CondEnt1WSI}
\end{align}
which we will now do.

We first provide an alternate expression for the left hand side
in~\eqref{eq:CondEnt1WSI}.
\begin{align}
H(\rvk| \rvw^N, \rvz^n) &= H(\rvk, \rvt^N |\rvw^N, \rvz^n)
-
H(\rvt^N | \rvk, \rvw^N, \rvz^n) \label{eq:detF0} \\
&= H(\rvt^N |\rvw^N, \rvz^n) -
H(\rvt^N | \rvk, \rvw^N, \rvz^n) \notag \\
&= H(\rvt^N,\Phi_\mrm{WZ} | \rvw^N, \rvz^n) - H(\rvt^N | \rvk, \rvw^N, \rvz^n)  \label{eq:detFt}\\
&= H(\Phi_\mrm{WZ} | \rvw^N, \rvz^n) + H(\rvt^N|\Phi_\mrm{WZ},\rvw^N) -
H(\rvt^N | \rvk, \rvw^N, \rvz^n) \label{eq:detFt2}
\end{align}
where~\eqref{eq:detFt} follows from the fact
that $\Phi_\mrm{WZ}$ is a deterministic function of $\rvt^N$,
while~\eqref{eq:detFt2} follows from the fact that $\rvt^N
\rightarrow (\rvw^N,\Phi_\mrm{WZ}) \rightarrow \rvz^n$ forms a Markov
chain. The right hand side in~\eqref{eq:CondEnt1WSI} is established by showing that
\begin{subequations}
\begin{align}
\frac{1}{n}H(\Phi_\mrm{WZ} | \rvw^N, \rvz^n) &\ge I(\rvx;\rvy|\rvz) +o_\eta(1) \label{eq:PhiEntWSI}\\
\frac{1}{n}H(\rvt^N|\Phi_\mrm{WZ},\rvw^N) &= \beta(I(\rvt;\rvv)-I(\rvt;\rvw))  + o_\eta(1)\label{eq:TcondPhiWSI}\\
\frac{1}{n} H(\rvt^N | \rvk, \rvw^N, \rvz^n) &= o_\eta(1).\label{eq:TcondKWZWSI}
\end{align}
\end{subequations} To interpret~\eqref{eq:PhiEntWSI}, recall that $\Phi_\mrm{WZ}$ is the message to the wiretap codebook.
The equivocation introduced by the wiretap codebook
$\frac{1}{n}H(\Phi_\mrm{WZ}|\rvz^n)$ equals $I(\rvx;\rvy|\rvz)$.
Eq.~\eqref{eq:PhiEntWSI} shows that if in addition to $\rvz^n$, the
eavesdropper has access to $\rvw^N$, a degraded source, the
equivocation still does not decrease (except for a negligible
amount). The intuition behind this claim is that since the bin index
$\Phi_\mrm{WZ}$ is almost independent of $\rvv^N$ (see
Lemma~\ref{lem:PhiVIndepCondn} below), it is also independent of
$\rvw^N$ due to the Markov condition.

Eq.~\eqref{eq:TcondPhiWSI} shows that the knowledge of $\rvw^N$
reduces the list of  $\rvt^N$ sequences in any bin from
$\exp(N(I(\rvt;\rvv)))$ to $\exp(N(I(\rvt;\rvv)- I(\rvt;\rvw)))$,
while~\eqref{eq:TcondKWZWSI} shows that for the code construction,
the eavesdropper, if revealed the secret-key, can decode $\rvt^N$
with high probability.

To establish~\eqref{eq:PhiEntWSI},
\begin{align}
\frac{1}{n}H(\Phi_\mrm{WZ} | \rvw^N, \rvz^n) &\ge
\frac{1}{n}H(\Phi_\mrm{WZ}| \rvz^n,\rvv^N)  \label{eq:PhiEntVWSI}\\
&= \frac{1}{n}H(\Phi_\mrm{WZ}|\rvz^n) - \frac{1}{n}I(\Phi_\mrm{WZ};\rvv^N|\rvz^n)\notag\\
&\ge I(\rvx;\rvy|\rvz) + o_\eta(1) - \frac{1}{n}I(\Phi_\mrm{WZ};\rvv^N|\rvz^n),\label{eq:PhiCondEquiv}\\
&\ge I(\rvx;\rvy|\rvz) + o_\eta(1) - \frac{1}{n}I(\Phi_\mrm{WZ};\rvv^N),\label{eq:PhiMarkov}
\end{align}
where~\eqref{eq:PhiEntVWSI} follows from the fact that $\rvw^N \rightarrow \rvv^N \rightarrow  \left(\Phi_\mrm{WZ}, \rvz^n\right)$,~\eqref{eq:PhiCondEquiv} from Lemma~\ref{lem:Uniformity} and~\eqref{eq:PhiMarkov} from the fact that $\rvv^N \rightarrow \Phi_\mrm{WZ} \rightarrow \rvz^n$ so that
\begin{equation}
\frac{1}{n}I(\Phi_\mrm{WZ};\rvv^N | \rvz^n) \le \frac{1}{n}I(\Phi_\mrm{WZ};\rvv^N).
\label{eq:PhiVindepNC}
\end{equation}
Thus we need to show the following.
\begin{lemma}
\begin{equation}
\frac{1}{n}I(\Phi_\mrm{WZ};\rvv^N) \le o_\eta(1).
\label{eq:Indep_desired_Condn}
\end{equation}\label{lem:PhiVIndepCondn}
\end{lemma}
\begin{new-proof}
From Lemma~\ref{lem:Uniformity} note that
$$\frac{1}{N}H(\Phi_\mrm{WZ}) = I(\rvt;\rvu)-I(\rvt;\rvv)+ o_\eta(1)$$
and hence we need to show that
$$\frac{1}{N}H(\Phi_\mrm{WZ}| \rvv^N) = I(\rvt;\rvu)-I(\rvt;\rvv)+ o_\eta(1)$$
as we do below.
\begin{align}
\frac{1}{N}H(\Phi_\mrm{WZ}| \rvv^N) &=\frac{1}{N}H(\Phi_\mrm{WZ},\rvt^N|\rvv^N) - \frac{1}{N}H(\rvt^N|\rvv^N, \Phi_\mrm{WZ}) \notag\\
&=\frac{1}{N}H(\rvt^N|\rvv^N) + o_\eta(1) \label{eq:tCondnV}\end{align}Where~\eqref{eq:tCondnV} follows
since each bin has $M_\mrm{WZ} = \exp\left(N(I(\rvt;\rvv)-\eta)\right)$ sequences, (from standard joint typicality arguments) we have that
\begin{equation}
\frac{1}{N}H(\rvt^N | \rvv^N, \Phi_\mrm{WZ}) = o_\eta(1).\label{eq:JointTypical}
\end{equation}
Finally  by substituting $\rva = \rvv$, $\rvb = \rvu$ and $\rvc = \rvt$ and $R =I(\rvt;\rvu) + \eta$,  in Lemma~\ref{lem:CondLemmaEq} in Appendix~\ref{app:condEntLem} we have that  $$\frac{1}{N}H(\rvt^N|\rvv^N) = I(\rvt;\rvu)-I(\rvt;\rvv) + o_\eta(1).$$
This completes the derivation of~\eqref{eq:Indep_desired_Condn}.

\end{new-proof}

To establish~\eqref{eq:TcondPhiWSI}, we again use Lemma~\ref{lem:CondLemmaEq} in Appendix~\ref{app:condEntLem}, with $\rva = \rvw$, $\rvb=\rvu$ and $\rvc = \rvt$ and $R = I(\rvt;\rvv) - \eta$. Finally, to establish~\eqref{eq:TcondKWZWSI}, we construct a decoder as in section~\ref{subsec:SecAnalysis} that searches for a sequence $\rvt_{kj}^N$ such that $\Phi_\mrm{WZ}(\rvt_{kj}^N) \in \cI_\mrm{x}$ and which is also jointly typical with $\rvw^N$. Since there are $\exp\{n(\beta I(\rvw;\rvt) + I(\rvx;\rvz)-\eta)\}$ sequences in the set, we can show along the same lines as in the proof of Lemma~\ref{lem:FanoLem} that $\rvt^N$ can be decoded with high probability given $(\rvk,\rvz^n,\rvw^N)$. The details will be omitted.

\subsection{Converse}
Suppose there is a sequences of $(n,N)$ codes that achieves a secret
key ($\rvk$) rate of $R$, and $\beta = N/n$. Then from Fano's
inequality,
$$H(\rvk | \rvy^n, \rvv^N) \le  n\eps_n,$$and from the secrecy constraint.
$$\frac{1}{n}I(\rvk; \rvz^n,\rvw^N )\le \eps_n.$$
Combining these inequalities, we have that,
\begin{align}
nR_\mrm{key} &\le  I(\rvk;\rvy^n,\rvv^N)-I(\rvk;\rvz^n,\rvw^N) + 2n\eps_n \notag\\
&\le  I(\rvk;\rvy^n,\rvv^N \mid \rvz^n,\rvw^N) + 2n\eps_n \notag\\
&\le h(\rvy^n\mid\rvz^n)+ h(\rvv^N\mid \rvw^N)-h(\rvy^n\mid \rvz^n,\rvw^N,\rvk)-h(\rvv^N\mid \rvy^n,\rvz^n,\rvw^N,\rvk)+ 2n\eps_n \notag\\
&\le h(\rvy^n\mid\rvz^n)+ h(\rvv^N\mid \rvw^N)-h(\rvy^n\mid \rvz^n,\rvw^N,\rvk, \rvx^n)-h(\rvv^N\mid \rvy^n,\rvz^n,\rvw^N,\rvk,)+ 2n\eps_n \notag\\
&=h(\rvy^n\mid\rvz^n)+ h(\rvv^N\mid \rvw^N)-h(\rvy^n\mid \rvz^n, \rvx^n)-h(\rvv^N\mid \rvy^n,\rvz^n,\rvw^N,\rvk,)+ 2n\eps_n \label{eq:chanIndep}\\
&\le \sum_{i=1}^n I(\rvx_i;\rvy_i \mid \rvz_i)+ h(\rvv^N\mid
\rvw^N)-h(\rvv^N | \rvy^n,\rvw^N,\rvk)+ 2n\eps_n \label{eq:MarkovZYW}\\
&\le n I(\rvx;\rvy \mid \rvz)+ h(\rvv^N\mid \rvw^N)-h(\rvv^N |
\rvy^n,\rvw^N,\rvk)+ 2n\eps_n\label{eq:RcondStep}
\end{align}where the~\eqref{eq:chanIndep}  follows from the fact that  $(\rvw^N,\rvk)\rightarrow (\rvz^n,\rvx^n)\rightarrow \rvy^n$, and~\eqref{eq:MarkovZYW} follows from the Markov condition $\rvz^n \rightarrow (\rvy^n,\rvw^n,\rvk)\rightarrow \rvv^N$ that holds for the degraded channel, while~\eqref{eq:RcondStep} follows from the fact that
$I(\rvx;\rvy|\rvz)$ is a concave function of $p_{\rvx_i}$ (see e.g.,~\cite[Appendix-I]{khistiTchamWornell:07}) and we select
$p_{\rvx}(\cdot) = \frac{1}{n}\sum_{i=1}^n p_{\rvx_i}(\cdot)$.
Now, let
$\rvt_i = (\rvk,\rvu_{i+1}^n\rvv^{i-1},\rvy^n)$, $J$ be a
random variable uniformly distributed over the set $[1,2,\ldots n]$
and $\rvt = (J,\rvk,\rvu_{J+1}^n\rvv^{J-1},\rvy^n)$ we have
that
\begin{align}
h(\rvv^N | \rvy^n,\rvw^N,\rvk)
&= \sum_{i=1}^N h(\rvv_i | \rvv^{i-1},\rvy^n, \rvw^N, \rvk) \notag\\
&\ge \sum_{i=1}^N h(\rvv_i | \rvv^{i-1}, \rvy^n, \rvw^N, \rvu_{i+1}^N,\rvk) \notag\\
&= \sum_{i=1}^N h(\rvv_i | \rvv^{i-1}, \rvy^n, \rvw_i,
\rvu_{i+1}^N,\rvk)  \label{eq:LongMarkov}\\
&= N \cdot h(\rvv_J | \rvt, \rvw_J) \notag
\end{align}
where we have used the fact that
$(\rvw^{i-1},\rvw_{i+1}^N)\rightarrow (\rvv^{i-1}, \rvy^n, \rvw_i,
\rvu_{i+1}^N,\rvk)\rightarrow \rvv_i$ which can be verified as follows
\begin{align}
& p\left(\rvv_i \mid \rvw_i, \rvw^{i-1},\rvw_{i+1}^N,  \rvv^{i-1}, \rvu_{i+1}^N,\rvy^n,
\rvk\right)\notag\\
=& \sum_{\rvu_i = u} p\left(\rvv_i \mid \rvw_i, \rvu_i = u, \rvw^{i-1},\rvw_{i+1}^N,  \rvv^{i-1}, \rvu_{i+1}^N,\rvy^n,
\rvk\right)p\left(\rvu_i = u \mid \rvw_i, \rvw^{i-1},\rvw_{i+1}^N,  \rvv^{i-1}, \rvu_{i+1}^N,\rvy^n,
\rvk \right)\notag\\
=& \sum_{\rvu_i = u} p\left(\rvv_i \mid \rvw_i, \rvu_i = u\right)p\left(\rvu_i = u \mid \rvw_i, \rvv^{i-1}, \rvu_{i+1}^N,\rvy^n,
\rvk \right)\label{eq:MarkovCondn2}\\
=& p\left(\rvv_i \mid \rvw_i,  \rvv^{i-1}, \rvu_{i+1}^N,\rvy^n,
\rvk\right)\notag,
\end{align}
where~\eqref{eq:MarkovCondn2} follows from the fact that since the sequence $\rvv^N$ is sampled i.i.d.\ , we have that
$$\rvv_i \rightarrow (\rvu_i,\rvw_i) \rightarrow (\rvw^{i-1},\rvw_{i+1}^N,  \rvv^{i-1}, \rvu_{i+1}^N,\rvy^n,
\rvk)$$
and since $\rvu \rightarrow \rvv \rightarrow \rvw$, it follows that
$$\rvu_i \rightarrow ( \rvv^{i-1}, \rvu_{i+1}^N,\rvy^n, \rvw_i,
\rvk) \rightarrow (\rvw^{i-1},\rvw_{i+1}^N).$$

Since, $\rvv_J$ and
$\rvw_J$ are both independent of $J$, we from~\eqref{eq:RcondStep}  that
$$R_\mrm{key} \le I(\rvx;\rvy|\rvz)+ \beta I(\rvt;\rvv|\rvw) + 2\eps_n.$$

Finally, using the steps between~\eqref{eq:RateConsStart}-\eqref{eq:RateConsStop} as in the converse for the case when
$\rvw =0$, we have that

\begin{equation}
I(\rvx;\rvy) \ge\beta(I(\rvt;\rvu)-I(\rvt;\rvv)),
\end{equation}
which completes the proof.

\section{Public discussion channel}
\label{sec:Discussion}

We establish the upper bound on the secret key capacity in the presence of interactive communication over a public discussion channel.

\begin{new-proof}

First from Fano's inequality we have the following,
\begin{align}
nR &= H(\rvk) \\
&= H(\rvk | \rvl) + I(\rvk; \rvl) \\
&\le n\eps_n +I(\rvk; \rvl)
\end{align}
where the last inequality follows from Fano's inequality. Also from the secrecy constraint we have that $$\frac{1}{n}I(\rvk; \phi^k, \psi^k, \rvz^n) \le \eps_n, $$ which results in the following
\begin{align}
nR &\le n\eps_n + I(\rvk; \rvl, \psi^k, \phi^k, \rvz^n)\\
&\le 2n \eps_n + I(\rvk; \rvl | \psi^k, \phi^k, \rvz^n)\\
&\le 2n \eps_n + I(\rvmx, \rvu^N; \rvm_\rvy, \rvv^N, \rvy^n | \psi^k, \phi^k, \rvz^n),\end{align}where the last step follows from the data-processing inequality since
$\rvk = K(\rvmx, \rvu^N, \psi^k) $ and  $\rvl =L(\rvm_\rvy,\rvv^N, \rvy^n,\phi^k)$.
\end{new-proof}

Using the chain rule, we have that
\begin{align}
&I(\rvmx, \rvu^N; \rvm_\rvy, \rvv^N, \rvy^n | \psi^k, \phi^k, \rvz^n)\\
&= I(\rvmx, \rvu^N; \rvm_\rvy, \rvv^N, \rvy^n, \psi^k, \phi^k, \rvz^n) - I(\rvmx, \rvu^N; \psi^k, \phi^k, \rvz^n)\\
&= I(\rvmx, \rvu^N; \rvm_\rvy, \rvv^N,\psi^{i_1-1}, \phi^{i_1-1}) + \sum_{j=1}^n F_j +G_j \notag\\
&\quad\quad - I(\rvmx, \rvu^N; \psi^{i_1-1}, \phi^{i_1-1}) - \sum_{j=1}^n \hat{F}_j + \hat{G}_j,\label{eq:chainRuleTerms}
\end{align} where for each $j = 1,2,\ldots, n$ we define  $F_j = I(\rvmx, \rvu^N; \rvy_j, \rvz_j|\rvm_\rvy, \rvv^N, \rvy^{j-1}, \rvz^{j-1}, \phi^{i_j-1}, \psi^{i_j-1} )$, $G_j = I(\rvmx, \rvu^N; \phi_{i_j +1}, \ldots, \phi_{i_{j+1}-1}, \psi_{i_j+1},\ldots, \psi_{i_{j+1}-1} | \rvm_\rvy,\rvv^N, \rvy^j, \rvz^j, \phi^{i_j-1}, \psi^{i_j-1} )$, and $\hat{F}_j = I(\rvmx, \rvu^N; \rvz_j | \rvz^{j-1}, \psi^{i_j-1},\phi^{i_j-1})$,   $\hat{G}_j = I(\rvmx, \rvu^N; \phi_{i_j +1}, \ldots, \phi_{i_{j+1}-1}, \psi_{i_j+1},\ldots, \psi_{i_{j+1}-1}| \rvz^j, \phi^{i_j-1}, \psi^{i_j-1})$.

We now bound the expression in~\eqref{eq:chainRuleTerms}. First note that
\begin{align*}
&I(\rvmx, \rvu^N; \rvm_\rvy, \rvv^N,\psi^{i_1-1}, \phi^{i_1-1})-I(\rvmx, \rvu^N; \psi^{i_1-1}, \phi^{i_1-1})\\&= I(\rvmx, \rvu^N; \rvm_\rvy, \rvv^N | \psi^{i_1-1}, \phi^{i_1-1})\\
&\le I(\rvmx,\rvu^N, \psi_{i_1-1}; \rvm_\rvy, \rvv^N | \psi^{i_1-2}, \phi^{i_1-1} )\\
&= I(\rvmx,\rvu^N ; \rvm_\rvy, \rvv^N | \psi^{i_1-2}, \phi^{i_1-1} )\\
&\le I(\rvmx,\rvu^N ; \rvm_\rvy, \rvv^N, \phi_{i_1-1} | \psi^{i_1-2}, \phi^{i_1-2} )\\
&= I(\rvmx,\rvu^N ; \rvm_\rvy, \rvv^N | \psi^{i_1-2}, \phi^{i_1-2} )
\end{align*}where the third and fifth step follow from the fact that $\psi_{i_1-1}= \Psi_{i_1-1}(\rvmx, \rvu^N, \phi^{i_1-2})$ and $\phi_{i_1-1} = \Phi_{i_1-1}(\rvm_\rvy, \rvv^N, \psi^{i_1-2})$. Recursively continuing we have that
\begin{equation}
I(\rvmx,\rvu^N; \rvm_\rvy, \rvv^N | \psi^{i_1-1}, \phi^{i_1-1} ) \le I(\rvmx, \rvu^N;\rvm_\rvy, \rvv^N) = I(\rvu^N;\rvv^N) = N I(\rvu;\rvv) \label{eq:recursiveUpperBound}
\end{equation}where we use the facts that $\rvmx \rightarrow \rvu^N\rightarrow \rvv^N\rightarrow \rvm_\rvy$ and that $(\rvu^N,\rvv^N)$ are discrete and memoryless.

Also note that
\begin{align}
& F_j - \hat{F}_j \\
&= \notag  I(\rvmx, \rvu^N; \rvy_j, \rvz_j|\rvm_\rvy, \rvv^N, \rvy^{j-1}, \rvz^{j-1}, \phi^{i_j-1}, \psi^{i_j-1} )- I(\rvmx, \rvu^N; \rvz_j | \rvz^{j-1}, \psi^{i_j-1},\phi^{i_j-1}) \notag\\
&= H(\rvy_j,\rvz_j | \rvm_\rvy, \rvv^N, \rvy^{j-1}, \rvz^{j-1}, \phi^{i_j-1}, \psi^{i_j-1}) - H(\rvy_j,\rvz_j | \rvm_\rvy, \rvv^N, \rvy^{j-1}, \rvz^{j-1}, \phi^{i_j-1}, \psi^{i_j-1},\rvmx, \rvu^N)\notag \\
&\quad - H(\rvz_j | \rvz^{j-1}, \psi^{i_j-1},\phi^{i_j-1}) + H(\rvz_j | \rvz^{j-1}, \psi^{i_j-1},\phi^{i_j-1}, \rvmx, \rvu^N) \notag\\
&= H(\rvy_j,\rvz_j | \rvm_\rvy, \rvv^N, \rvy^{j-1}, \rvz^{j-1}, \phi^{i_j-1}, \psi^{i_j-1}) - H(\rvy_j,\rvz_j | \rvx_j)  - H(\rvz_j | \rvz^{j-1}, \psi^{i_j-1},\phi^{i_j-1}) + H(\rvz_j | \rvx_j) \label{eq:condnXval}\\
&\le H(\rvy_j | \rvz^j, \psi^{i_j-1},\phi^{i_j-1}) - H(\rvy_j|\rvz_j,\rvx_j) \notag\\
&\le I(\rvx_j;\rvy_j|\rvz_j)\label{eq:recursiveFBound},
\end{align}
where~\eqref{eq:condnXval} follows from the fact that $\rvx_j = X_j(\rvmx, \rvu^N, \psi^{i_j-1})$ and that since the channel is memoryless $(\rvmx,\rvm_\rvy,\rvu^N,\rvv^N, \phi^{i_j-1}, \psi^{i_j-1}, \rvy^{j-1},\rvz^{j-1}) \rightarrow \rvx_j \rightarrow (\rvy_j,\rvz_j) $ holds. The last two steps follow from the fact that conditioning reduces entropy.

Finally to  upper bound $G_j - \hat{G}_j$,

\begin{align*}
&G_j - \hat{G}_j \\
&= I(\rvmx, \rvu^N; \phi_{i_j +1}, \ldots, \phi_{i_{j+1}-1}, \psi_{i_j+1},\ldots, \psi_{i_{j+1}-1} | \rvm_\rvy,\rvv^N, \rvy^j, \rvz^j, \phi^{i_j-1}, \psi^{i_j-1} )  \\
&\quad- I(\rvmx, \rvu^N; \phi_{i_j +1}, \ldots, \phi_{i_{j+1}-1}, \psi_{i_j+1},\ldots, \psi_{i_{j+1}-1}| \rvz^j, \phi^{i_j-1}, \psi^{i_j-1})\\
&= I(\rvmx,\rvu^N; \rvm_\rvy, \rvv^N, \rvy^j, \phi_{i_j +1}, \ldots, \phi_{i_{j+1}-1}, \psi_{i_j+1},\ldots, \psi_{i_{j+1}-1}| \rvz^j, \phi^{i_j-1}, \psi^{i_j-1}  ) \\
&- I(\rvmx,\rvu^N; \rvm_\rvy, \rvv^N, \rvy^j|\rvz^j, \phi^{i_j-1}, \psi^{i_j-1}  )\!-\!I(\rvmx, \rvu^N\!;\! \phi_{i_j +1}, \ldots, \phi_{i_{j+1}-1}, \psi_{i_j+1},\ldots, \psi_{i_{j+1}-1}| \rvz^j, \phi^{i_j-1},\!\!\psi^{i_j-1}\!\!)\\
&= I(\rvmx,\rvu^N;\rvm_\rvy,\rvv^N,\rvy^j | \phi^{i_{j+1}-1},\psi^{i_{j+1}-1},\rvz^j) - I(\rvmx,\rvu^N;\rvm_\rvy,\rvv^N,\rvy^j | \phi^{i_{j}-1},\psi^{i_{j}-1},\rvz^j)
\end{align*}
Furthermore since $\phi_{i_{j+1}-1} = \Phi_{i_{j+1}-1}(\rvmx, \rvu^N, \psi^{i_{j+1}-2})$ and $\psi_{i_{j+1}-1} = \Psi_{i_{j+1}-1}(\rvm_\rvy, \rvv^N, \phi^{i_{j+1}-2})$ we have that
\begin{align*}
&\quad I(\rvmx,\rvu^N;\rvm_\rvy,\rvv^N,\rvy^j | \phi^{i_{j+1}-1},\psi^{i_{j+1}-1},\rvz^j) \\
&\le I(\rvmx,\rvu^N, \phi_{i_{j+1}-1};\rvm_\rvy,\rvv^N,\rvy^j | \phi^{i_{j+1}-2},\psi^{i_{j+1}-1},\rvz^j) \\
&= I(\rvmx,\rvu^N ;\rvm_\rvy,\rvv^N,\rvy^j | \phi^{i_{j+1}-2},\psi^{i_{j+1}-1},\rvz^j) \\
&\le I(\rvmx,\rvu^N ;\rvm_\rvy,\rvv^N,\rvy^j,\psi_{i_{j+1}-1} | \phi^{i_{j+1}-2},\psi^{i_{j+1}-2},\rvz^j) \\
&= I(\rvmx,\rvu^N ;\rvm_\rvy,\rvv^N,\rvy^j, |
\phi^{i_{j+1}-2},\psi^{i_{j+1}-2},\rvz^j)
\end{align*}
Continuing this process we have that
$$I(\rvmx,\rvu^N;\rvm_\rvy,\rvv^N,\rvy^j | \phi^{i_{j+1}-1},\psi^{i_{j+1}-1},\rvz^j) \le  I(\rvmx,\rvu^N;\rvm_\rvy,\rvv^N,\rvy^j | \phi^{i_{j}-1},\psi^{i_{j}-1},\rvz^j) $$ and thus
\begin{equation}
G_j - \hat{G}_j \le 0. \label{eq:recursiveGineq}
\end{equation}
Substituting~\eqref{eq:recursiveUpperBound},~\eqref{eq:recursiveFBound} and~\eqref{eq:recursiveGineq} into~\eqref{eq:chainRuleTerms} we have that
\begin{align}
nR &\le \sum_{j=1}^n I(\rvx_j;\rvy_j|\rvz_j) + N I(\rvu;\rvv) + 2n\eps_n \\
&\le \max_{p_\rvx}  n I(\rvx;\rvy|\rvz) +  N I(\rvu;\rvv) + 2n\eps_n
\end{align} thus yielding the stated upper bound.

\section{Conclusions}
\label{sec:concl} In this paper we introduced a secret-key agreement
technique that harnesses uncertainties from both sources and
channels. Applications of sensor networks and biometric systems
motivated this setup.

We first consider the case when the legitimate terminals observe a
pair of correlated sources and communicate over a wiretap channel
for generating secret keys. The secret-key capacity is bounded by
establishing upper and lower bounds. The lower bound is established
by providing a coding theorem that combines ideas from source and
channel coding.  Its  optimality is established when the wiretap
channel consists of parallel, independent and degraded channels. The
lower bound in general involves us to operate at a point on the
wiretap channel that balances the contribution of source and channel
contributions and this illustrated for the Gaussian channels.

In addition we also establish the capacity when the wiretapper  has
access to a source sequence which is a degraded version of the
source sequence of the legitimate receiver. Furthermore the case
when a public discussion channel is available for interactive
communication is also studied and an upper bound on the secret-key
capacity is provided. For the practically important case, when the
wiretap channel consists of ``independent noise" for the legitimate
receiver and the discussion channel allows us to separately generate
keys from source and channel components without loss of optimality.

In terms of future work, there can be many fruitful avenues to
explore for  secret-key distillation in a joint-source-channel
setup. One can consider multi-user extensions of the secret-key
generation problem along the lines of~\cite{csiszarNarayan:04} and
also consider more sophisticated  channel models such as the
compound wiretap channels,  MIMO wiretap channels and wiretap
channels with feedback and/or side information. Connections of this
setup to wireless channels, biometric systems and other applications
can also be interesting.

\appendices

\section{Extension of Lemma~\ref{lem:achievRate} to  general $(\rva,\rvb)$}

\label{app:GenAchiev}
We extend the coding theorem in section~\ref{sec:AchievMain} for Lemma~\ref{lem:achievRate} to the case of general $(\rva,\rvb)$.

We focus on the case when  $\rva = \rvx$. The general case then follows by further considering the auxiliary channel $\rva \rightarrow \rvx$, sampling the codewords from the typical set $T_\rva^n$ and then passing each symbol of $\rva^n$ through an auxiliary channel $p_{\rvx|\rva}(\cdot)$.

Our extension involves using a superposition code as discussed below. Let us define  $R_\mrm{a} = I(\rvx;\rvy|\rvb)$ and $R_\mrm{b} = I(\rvb;\rvy)$. Since $\rvb\rightarrow\rvx \rightarrow \rvy$, we have that $R_\mrm{b} +R_\mrm{a}= I(\rvx;\rvy)$.   We first generate a codebook $\cC_\mrm{b}$ with $N_\mrm{b}=\exp\left(n (R_\mrm{b}-\del_b)\right)$ sequences sampled uniformly from the set $T_{\rvb}^n$.
For each sequence $\rvb_i^n \in \cC_\mrm{b}$, we generate a codebook $\cC_\mrm{a}(\rvb_i^n)$ by selecting $N_\mrm{a} = \exp(n(I(\rvx;\rvy|\rvb)-\del_a))$ sequences uniformly at random from the set $T^n_{\rvx,\rvb}(\rvb_i^n)$.

Select $\del_{a} > 0$ and $\del_\mrm{b} > 0$ as arbitrary constants such that $\del_\mrm{a} +\del_\mrm{b} = \del$, which satisfies~\eqref{eq:tightRateCons}. Note that we have $N_\mrm{WZ} = N_\mrm{a}\cdot N_\mrm{b}$. We define an encoding functions: $\Phi_\mrm{WZ,\rvb}: \{1,2,\ldots, N_\mrm{b}\} \rightarrow \cC_b$  and $\Phi_\mrm{WZ,\rva}^i: \{1,2,\ldots, N_\mrm{a}\} \rightarrow \cC_a(\rvb_i^n)$ as a mapping from the messages to respective codewords in the codebooks.

The construction of the Wyner-Ziv codebook and the secret-key codebook is via random partitioning  along the lines in section~\ref{subsec:codebookCons} --- the constants $M_\mrm{WZ}$ and $N_\mrm{WZ}$ are as given in~\eqref{eq:MWZ} and~\eqref{eq:NWZ} respectively while\begin{subequations}
 \begin{align} & M_\mrm{SK} = \exp\left(n(I(\rvb;\rvy)+ I(\rvx;\rvz|\rvb)-\del)\right),\label{eq:MSK_Sup}\\
& N_\mrm{SK} = \exp\left(n(\beta I(\rvt;\rvv)+ I(\rvx;\rvy|\rvb)-I(\rvx;\rvz|\rvb)-\del)\right).\label{eq:NSK_Sup}
\end{align}
\end{subequations}

The encoding function is defined as follows:
given a sequence $\rvu^N$, as in section~\ref{subsec:Encoding}, a jointly typical sequence  $\rvt^N \in \cT$ is selected and the  bin index and secret-key are computed via the mappings  $\Phi_\mrm{WZ}(\rvt^N)$ and $\Phi_\mrm{SK}(\rvt^N)$ respectively in Def.~\ref{def:encoding_funcs}. The bin index is split into two indices $\Phi_\mrm{a} \in\{1,2,\ldots, N_\mrm{a}\}$ and $\Phi_\mrm{b} \in \{1,\ldots, N_\mrm{b}\}$, which  form messages for the channel codebooks constructed above and the resulting sequence $\rvx^n$ is transmitted.

The decoder upon observing $\rvy^n$ searches for sequences $\rvb_i^n \in \cC_\mrm{b}$ and $\rvx^n \in \cC_\mrm{a}(\rvb_i^n)$ that are jointly typical i.e., $(\rvy^n,\rvx^n,\rvb_i^n) \in T_{\rvy,\rvx,\rvb,\eta}^n$. By our choice of $N_\mrm{b}$  and $N_\mrm{a}$ this succeeds with high probability. It then reconstructs the bin index $\Phi_\mrm{WZ}$ and searches for a sequence $\rvt^N \in\cT$ that lies in this bin and is jointly typical with $\rvv^N$. As in section~\ref{subsec:Decoding}, this step succeeds with high probability. The secret-key is then computed as $\rvhk = \Phi_\mrm{SK}(\rvt^N)$.

We need to show the secrecy condition that \begin{equation}
\frac{1}{n} H(\rvk | \rvz^n) = \{I(\rvx;\rvy|\rvb)-I(\rvx;\rvz|\rvb)\} + \beta I(\rvt;\rvv) + o_\eta(1).\label{eq:SecrecyCondnFull}
\end{equation}

By expressing $H(\rvk|\rvz^n)$ as in~\eqref{eq:VTPhMarkov} in section~\ref{subsubsec:EquivAnalysis} \begin{equation}
H(\rvk|\rvz^n) = H(\Phi_\mrm{WZ}|\rvz^n) + H(\rvt^N | \Phi_\mrm{WZ}) - H(\rvt^N | \rvk, \rvz^n). \label{eq:SecCondnd2}
\end{equation}For the superposition codebook, since $\Phi_\mrm{WZ}$ is the transmitted message we have from~\cite{csiszarKorner}
\begin{equation}
\frac{1}{n}H(\Phi_\mrm{WZ}|\rvz^n) =  I(\rvx;\rvy|\rvb)-I(\rvx;\rvz|\rvb) + o_\eta(1),
\label{eq:Sec1}
\end{equation}and from~\eqref{eq:PhitWZUnif} in  Lemma~\ref{lem:Uniformity},
\begin{equation}
\frac{1}{N}H(\rvt^N | \Phi_\mrm{WZ} ) = I(\rvt;\rvv) + o_\eta(1).\label{eq:Sec2}
\end{equation}To show that
\begin{equation}\frac{1}{N}H(\rvt^N|\rvz^n,\rvk) = o_\eta(1)\label{eq:Sec3}\end{equation}
we use a decoder analogous to that in the proof of Lemma~\ref{lem:FanoLem} in Section~\ref{subsec:SecAnalysis}. Upon observing $\rvz^n$, the decoder searches for a sequence $\rvb_i^n \in \cC_\mrm{b}$ that is jointly typical. This event succeeds with high probability since $I(\rvb;\rvz) \ge I(\rvb;\rvy) = R_\mrm{b}$. Let the set of conditionally typical sequences $\rvx^n$ be
\begin{equation}
\cI_\mrm{x} = \{ j | \rvx_j^n \in \cC_\mrm{b}(\rvb_i^n) , (\rvx_j^n,\rvz^n)\in T_{x,z,\eta}^n\}.
\end{equation}
The eavesdropper searches for all sequences $\rvt_{kj,\mrm{SK}}^N$ such that $\Phi_\mrm{a}(\rvt_{kj,\mrm{SK}}^N) \in \cI_\mrm{x}$ and $\Phi_\mrm{b}(\rvt_{kj,\mrm{SK}}^N)=i $. Since the number of sequences $\rvt_{kj,\mrm{SK}}^N$ is $M_\mrm{SK}=\exp\left(n(I(\rvx;\rvz|\rvb)+ I(\rvb;\rvy)-\del)\right)$, along the lines of Lemma~\ref{lem:FanoLem}, it follows that the codeword sequence is decoded with high probability.

Note that~\eqref{eq:SecrecyCondnFull} follows from~\eqref{eq:SecCondnd2},~\eqref{eq:Sec1},~\eqref{eq:Sec2} and~\eqref{eq:Sec3}.

\section{Conditional Entropy Lemma}
\label{app:condEntLem}
\begin{lemma}
Suppose that the random variables $\rva$, $\rvb$, and $\rvc$ are
finite valued with a joint distribution $p_{\rva,\rvb,\rvc}(\cdot)$
that satisfies $\rva\rightarrow \rvb\rightarrow \rvc$. Suppose that
a set $\cC_c$ is selected by drawing  $\exp(NR)$ sequences
$\{c_i^N\}$ uniformly and at random from the set of typical
sequences $T_\rvc^N$ where $R < H(\rvc)$.  Suppose that the pair of
length-$N$ sequences $(\rva^N,\rvb^N)$  are drawn i.i.d. from the
distribution $p_{\rva,\rvb}$ and a sequence $c_i^N \in \cC_c$ is
selected uniformly at random from the set of all possible sequences
such that $(c_i^N,b^N)\in T_{\rvc\rvb, \eta}^N$. Then for $R
> I(\rvc;\rva)$, we have that
\begin{equation}
\label{eq:CondLemmaEq}
\frac{1}{N}H(\rvc_i^N|\rva^N) = R - I(\rvc;\rva) + o_\eta(1),
\end{equation}where  the term $o_\eta(1)$ vanishes to zero as $N\rightarrow \infty$ and  $\eta \rightarrow 0$.
\label{lem:CondLemmaEq}
\end{lemma}
\begin{new-proof}
From~\eqref{eq:typeSetProb}, for all pair of sequences $(\rva^N, \rvb^N)$, except a set whose  probability is $o_\eta(1)$, we have that $(a^N,b^N)\in T_{\rva\rvb,\eta}^N$.  For each such typical pair, since $\rva \rightarrow \rvb \rightarrow \rvc$ and $(\rvb^N, \rvc^N_i) \in T_{\rvb\rvc,\eta}^N$ from the Markov Lemma it follows that $(a^N, c^N_i) \in T_{\rva\rvc,\eta}^N$.

To establish~\eqref{eq:CondLemmaEq} it suffices to show that for all sequences $a^N \in T_{\rva,\eta}^N$, except a set whose probability is at most $o_\eta(1)$
\begin{equation}
\Pr(\rvc^N = c_i^N | \rva^N=a^N) = \exp({-N(R-I(\rvc;\rva)+ o_\eta(1))}).\label{eq:CondProbEq}
\end{equation}
The expression in~\eqref{eq:CondLemmaEq} then immediately follows by
due to the continuity of the $\log$ function.  To
establish~\eqref{eq:CondProbEq},
\begin{equation}\label{eq:bayes}\Pr(\rvc^N = c_i^N | \rva^N = a^N) = \frac{p(a^N|c_i^N)\Pr(\rvc^N = c_i^n)}{p(a^N)}.\end{equation}
From property~\eqref{eq:typeProb} of typical sequences $p(a^N) =
\exp(-N(H(\rva)+ o_\eta(1)))$, $p(a^N|c_i^N) = \exp(-N(H(\rva|\rvc)+
o_\eta(1)))$ and since the sequence $\rvc^N$ is uniformly selected
from $2^{nR}$ sequences, we have that $\Pr(\rvc^N = c_i^N) =
\exp(-NR)$. Substituting these quantities in~\eqref{eq:bayes}
establishes~\eqref{eq:CondProbEq}.\end{new-proof}

\bibliographystyle{IEEEtranS}

\end{document}